\documentclass[journal]{IEEEtran}
\usepackage{textcomp}
\usepackage{url}
\usepackage{verbatim}
\usepackage{etoolbox}
 \usepackage{amsmath,amssymb}
 \usepackage{graphicx,graphics,color,psfrag}
 \usepackage{balance}
 
 \allowdisplaybreaks[4]
 \usepackage{algorithm}
 \usepackage{accents}
 \usepackage{amsthm}
 \usepackage{bm}
 \usepackage{algorithmic}
 \usepackage[english]{babel}
 \usepackage{multirow}
 \usepackage{enumerate}
 \usepackage{cases}
 \usepackage{stfloats}
 \usepackage{dsfont}
 \usepackage{color,soul}
 \usepackage{amsfonts}
 \usepackage{graphicx,amsmath,amssymb}
 \usepackage{fancyhdr}
 \usepackage{hhline}
 \usepackage{graphicx,graphics}
 \usepackage{array,color}
 \usepackage{amsmath}
\usepackage{float}
\usepackage{amssymb}
\usepackage{amsmath}
\usepackage{bbm}
\usepackage{amsthm}
\usepackage{amsfonts}
\usepackage{graphicx}
\usepackage{algorithm}
\usepackage{algorithmic}
\usepackage{epstopdf}
\usepackage{cite}
\usepackage{amsmath,bm}
\usepackage{subfigure}
\usepackage{graphicx}
\usepackage{color}
\usepackage{graphicx}
\usepackage{calc}
\usepackage{diagbox}

\begin{document}
\bibliographystyle{IEEEtran}

\title{Knowledge Distillation and Training Balance for Heterogeneous Decentralized Multi-Modal Learning over Wireless Networks \thanks{B. Yin and Z. Chen are with Cooperative Medianet Innovation Center and Shanghai Key Laboratory of Digital Media Processing and Transmission, Shanghai Jiao Tong University, Shanghai, China (e-mail: \{yinbsh, zhiyongchen\}@sjtu.edu.cn). M. Tao is with Department of Electronic Engineering, Shanghai Jiao Tong University, China (e-mail: mxtao@sjtu.edu.cn). \emph{(Corresponding author: Zhiyong Chen, Meixia Tao)}}}

\author{Benshun Yin, Zhiyong Chen, \emph{Senior Member, IEEE}, and Meixia Tao, \emph{Fellow, IEEE}}
%\author{
%\authorblockN{XXX} \\
%\authorblockA{Shanghai Jiao Tong University, China} \\
%Email: xxx@sjtu.edu.cn \\
%Tel: +86 - 10086}
\maketitle

%%%%%%%%%%%%%%%%%%%%%%%%%%%%%%%%%%%%%%%
\begin{abstract}
Decentralized learning is widely employed for collaboratively training models using distributed data over wireless networks. Existing decentralized learning methods primarily focus on training single-modal networks. For the decentralized multi-modal learning (DMML), the modality heterogeneity and the non-independent and non-identically distributed (non-IID) data across devices make it difficult for the training model to capture the correlated features across different modalities. Moreover, modality competition can result in training imbalance among different modalities, which can significantly impact the performance of DMML. To improve the training performance in the presence of non-IID data and modality heterogeneity, we propose a novel DMML with knowledge distillation (DMML-KD) framework, which decomposes the extracted feature into the modality-common and the modality-specific components. In the proposed DMML-KD, a generator is applied to learn the global conditional distribution of the modality-common features, thereby guiding the modality-common features of different devices towards the same distribution. Meanwhile, we propose to decrease the number of local iterations for the modalities with fast training speed in DMML-KD to address the imbalanced training. We design a balance metric based on the parameter variation to evaluate the training speed of different modalities in DMML-KD. Using this metric, we optimize the number of local iterations for different modalities on each device under the constraint of remaining energy on devices. Experimental results demonstrate that the proposed DMML-KD with training balance can effectively improve the training performance of DMML.
\end{abstract}

\begin{IEEEkeywords}
Decentralized learning, multi-modal learning, knowledge distillation, training balance
\end{IEEEkeywords}

\section{Introduction}
Machine learning, exemplified by deep neural networks, has achieved significant breakthroughs in various domains \cite{lecun2015deep}, such as object detection and image classification. With the rapid expansion of the Internet of Things (IoT), a substantial volume of data is being generated by distributed IoT devices, such as wireless security cameras. To harness the vast amount of data for intelligent applications, edge learning \cite{Deng2020,zhu2020toward,wang2020conv,federated,fed_split,designfed, chenjsac,dec_d2d,fed_d2d,data_nas} is emerging as a promising research field. 

Federated learning (FL) \cite{federated,designfed,fed_split} is a edge learning algorithm that can collaboratively train a neural network with the local dataset of wireless devices. Compared with uploading large amounts of raw data to the edge server for centralized learning, FL can protect the data privacy of users. However, certain problems persist in FL systems. Firstly, relying on an edge server for model aggregation may introduce a single point of failure, where an attack on the server can disrupt the entire FL system \cite{fed_block}. Secondly, considering the vast scalability of the edge computing networks, a single edge server cannot effectively aggregate all the updates from a large number of devices. Thirdly, transmitting the model parameters to the edge server is susceptible to security vulnerabilities in realistic wireless networks, such as malicious threats that can tamper with or steal local update information\cite{fed_attack}. Consequently, FL may be not applicable in some scenarios, making it necessary to apply the decentralized learning \cite{chenjsac,dec_d2d,fed_d2d} approach that eliminates the reliance on a server. 

On the other hand, with the rapid development of sensor technology, the physical environment can be detected from various modalities. Similar to how humans perceive the world through multiple senses, multi-modal sensory data can provide comprehensive information from multiple aspects. Multi-modal learning \cite{Gao_2020_CVPR,MISA} has become a recent trend in machine learning, which can obtain performance improvement compared to single-modal learning by integrating multi-modal data. For example, human action recognition \cite{Gao_2020_CVPR} and multi-modal sentiment analysis \cite{MISA} can extract multi-modal feature representation from different modalities including audio and vision, thereby improving prediction performance. Multiple sensor data including LiDAR, GPS and camera images can be used to speed up the sector selection in the mmWave band for vehicular mobility scenarios\cite{fed_mmwave}. 

Existing decentralized learning approaches \cite{chenjsac,dec_d2d,fed_d2d} have not considered the characteristics of multi-modal learning. Compared to decentralized single-modal learning, the efficient training of decentralized multi-modal learning (DMML) can not be achieved by simply changing the training model from single-modal to multi-modal. Besides the challenge of non-independent and identically distributed (non-IID) data among different devices, the DMML systems also face the issue of modality heterogeneity. Due to variations in detection environments and device types, the modalities of data collected by different wireless devices may differ. For example, some automobiles may be equipped with visual sensors but lack radar sensors. In dynamic environments, certain sensors may not always be available, leading to the absence of certain modalities in the collected data. Therefore, the data on different devices may exhibit different combinations of modalities. While many decentralized learning approaches have been proposed to address the challenge of non-IID data across devices, they may have a potential limitation, as they typically require all participating devices to have the same modality. In multi-modal learning, different neural networks are typically used to extract feature representations for different modalities. For instance, the transformer model \cite{transformer} can be used to process textual data, while convolutional neural networks are employed for image data processing. Modality heterogeneity implies that the models trained on different devices are different. Besides, jointly training a multi-modal network with multiple input modalities enables the network to learn correlated features across different modalities. However, the absence of certain modalities for some devices increases the difficulty of learning the cross-modal features in DMML.

\begin{figure}[t]
	\centering
	\includegraphics[width=8cm]{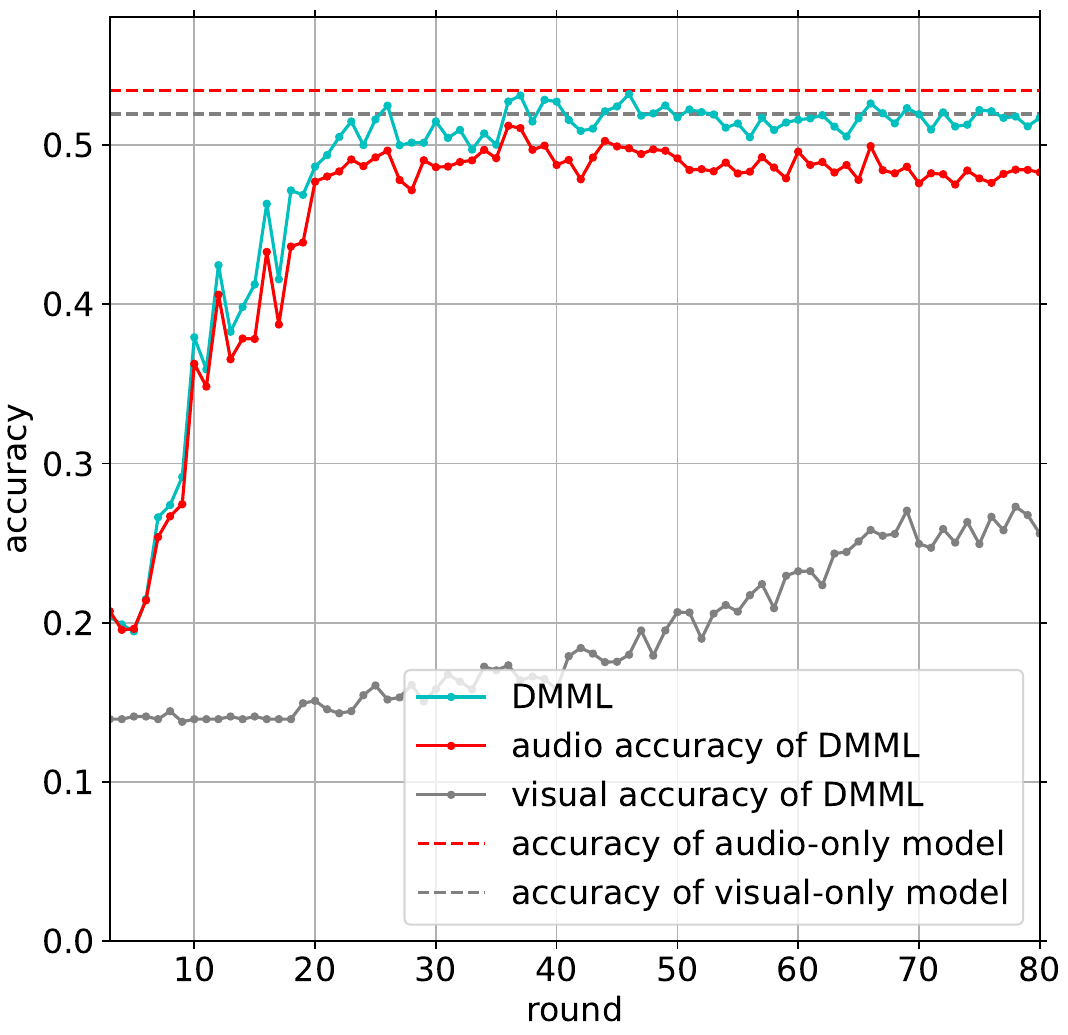}
	\caption{Training imbalance of DMML with CREMA-D dataset \cite{cremad}.}
	\label{imbalance}
\end{figure}

In addition to non-IID and modality heterogeneity, another challenge of DMML is the imbalanced training of different modalities \cite{modal_comp,HGB,OGM}. The performance of the individually trained single-modal model may surpass that of the jointly trained multi-modal network \cite{HGB}. In \cite{modal_comp}, it is theoretically proved that when training a single-modal network, features can be sufficiently learned, resulting in good performance. However, modality competition exists when training a multi-modal network. Only a subset of modalities' feature extraction networks can adequately learn the feature representations, leading to imbalanced training across different modalities. As shown in Fig. \ref{imbalance}, we estimate the accuracy of the two modalities under the joint training. The training of the visual modality is insufficient, which can not achieve the accuracy obtained from training the visual-only network. In DMML systems, the local training of a device with multiple modalities is inherently imbalanced. Due to the modality heterogeneity and non-IID data, the training imbalance of different modalities varies across devices, which consequently exacerbates the imbalance after the parameter aggregation among devices. These issues motivate us to propose the knowledge distillation and training balance method to improve the learning performance of DMML over wireless networks.

%2p decentralized learning + multi modal learning (non-iid, modal lack) + unbalance fig 

\subsection{Related Work}
Several recent works \cite{fed_mmwave,fedHGB,FedMSplit,contra_rep,fed_HAR} have considered multi-modal learning in federated learning systems. In \cite{fed_mmwave}, it assumes that all vehicles have the same sensors, ensuring identical modalities across devices. To compensate for missing sensor data during model inference, \cite{fed_mmwave} utilizes previous data from the same sensor. The training scheme of federated multi-modal learning is improved in \cite{FedMSplit,contra_rep,fed_HAR} to address the issue of modality heterogeneity. In \cite{FedMSplit}, a dynamic and multi-view graph structure corresponding to the devices is applied on the edge server to automatically capture the relationships among devices, which is utilized to achieve correlated local model updates among devices. Using the global model and the local model's outputs with respect to a public dataset, an inter-modal contrastive objective is designed in \cite{contra_rep} to complement for the absent modality in some users, and an intra-modal contrastive objective is proposed to regularize models to converge towards the global consensus. To learn the cross-modal features, the modality-agnostic features and the modality-specific features are extracted from the data of each modality separately in \cite{fed_HAR}, and a modality discriminator is employed to guide the training of the feature extraction networks. However, the algorithm designed specifically for federated learning cannot be directly applied to decentralized learning due to its reliance on the server.

Existing decentralized learning methods \cite{chenjsac,dec_d2d,fed_d2d} focus on the single-modal learning, such as training single-modal image classifiers. Many works have addressed the non-IID data problem in decentralized learning \cite{noniid_Quag,cross_grad,learn_coll}. The authors in \cite{noniid_Quag} conduct detailed experiments to demonstrate the impact of non-IID data on the performance of decentralized learning. To deal with the non-IID data, a cross-gradient aggregation algorithmic framework is proposed in \cite{cross_grad}. In \cite{learn_coll}, the mixing-weights of devices are dynamically updated to improve the personalized model for each device’s task. Besides, some works have studied the impact of topology on decentralized learning \cite{role_topo,exper_driven,EE_topo,Unified_topo}. The authors in \cite{role_topo} theoretically quantify how the topology affects the convergence of decentralized learning. In \cite{exper_driven}, an integer programming problem is formulated to find the best suitable topology at each training round. In \cite{EE_topo}, the topology of devices is optimized to save the transmission energy consumption. It is proved in \cite{Unified_topo} that the condition for achieving consensus in decentralized learning can be relaxed to the requirement that the merged topology of multiple training rounds is a connected graph. 

Several methods \cite{HGB,fedHGB,OGM,balan_LR} have been proposed to mitigate the challenge of imbalanced multi-modal training. In \cite{HGB}, the overfitting and generalization behaviors of the training multi-modal model are quantified by the overfitting-to-generalization ratio, which is minimized to determine optimal gradient blending weights of different modalities. Based on the assumption of convex loss functions, the gradient blending is extended from centralized learning to federated learning in \cite{fedHGB}. The authors in \cite{OGM} propose to adaptively control the optimization of each modality based on the discrepancy of their contribution towards the learning objective, thereby addressing the training imbalance of different modalities. In \cite{balan_LR}, the learning processes of single-modal and multi-modal networks are decoupled, and the learning rates of different modalities are dynamically balanced using an adaptive tracking factor. Different from the aforementioned approaches, we propose a training balance metric specific designed for decentralized learning in this paper.
\begin{figure}[t]
	\centering
	\includegraphics[width=8.5cm]{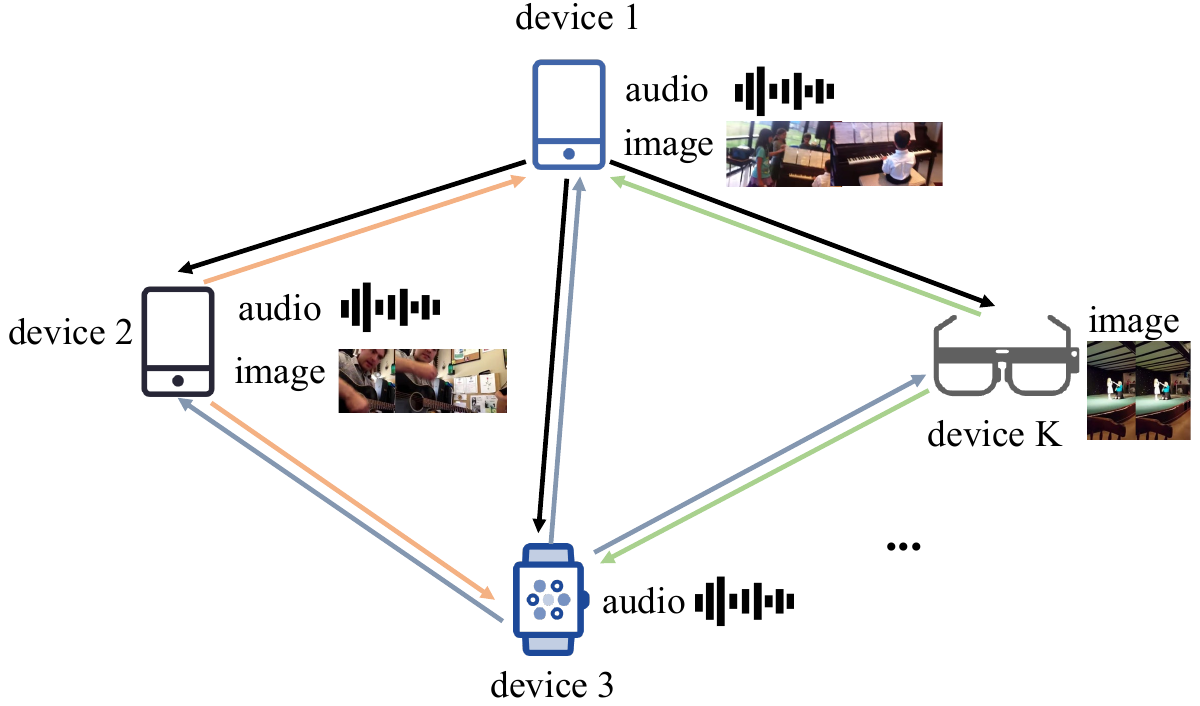}
	\caption{A decentralized multi-modal learning system.}
	\label{system}
\end{figure}

\subsection{Contributions and Outline}
Motivated by the above, we propose the decentralized multi-modal learning with knowledge distillation (DMML-KD) framework to deal with the problem of non-IID data and modality heterogeneity. Furthermore, we design a training balance method to improve the learning performance of decentralized multi-modal learning over wireless networks. The main contributions are summarized as follows: 

\begin{itemize}
\item We propose the DMML-KD framework to improve the training performance under the non-IID data and modality heterogeneity. We firstly decompose the extracted feature into the modality-common and the modality-specific components. Then we use the generator to learn the global conditional distribution of the modality-common features. The devices can utilize the generator to guide its modality-common features towards the global distribution, thereby potentially capturing correlated features across different modalities.

\item We develop a method to control the number of local iterations for different modalities to deal with the imbalanced training of different modalities. Compared to adjusting the learning rates \cite{OGM,balan_LR} of different modalities, the proposed method can reduce the computational burden of devices, thereby reducing the energy consumption.

\item We design a performance metric based on the parameter variation to evaluate the training speed of different modalities. Based on the metric, we optimize the number of iterations for different modalities. Experimental results show that the proposed DMML-KD framework with training balance can effectively improve the training performance of DMML.
\end{itemize}

The rest of this paper is organized as follows. The system model and the proposed DMML-KD framework are introduced in Section \ref{sys_model}. The training balance method and the optimization problem formulation are presented in Section \ref{prob}. The detail of the proposed training balance algorithm is introduced in Sections \ref{alg}. Finally, extensive experimental results are presented in Section \ref{sim}, and conclusions are drawn in Section \ref{con}.

%%%%%%%%%%%%%%%%%%%%%%%%%%%%%%%%%%%%%%%
\section{System Model and DMML-KD framework}
In this section, we introduce a decentralized multi-model learning system, focusing on the design of the system model and loss functions. We then develop the DMML-KD framework.
\label{sys_model}
%\begin{figure}[t]
%	\centering
%	\subfigure[]{\includegraphics[width=2.4cm]{fig/NAS.eps}}
%	\subfigure[]{\includegraphics[width=3.5cm]{fig/node.eps}}
%	\caption{(a) Overall architecture of the NAS space. (b) An example of a normal cell.}
%	\label{NAS}
%\end{figure}

\begin{figure}[t]
	\centering
	\includegraphics[width=5.7cm]{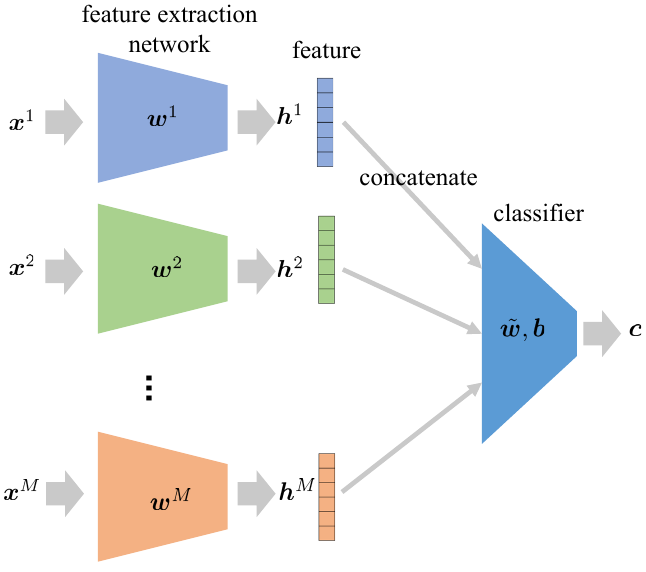}
	\caption{A multi-modal neural network with the concatenation fusion method.}
	\label{model}
\end{figure}
\subsection{System Model}
We consider a decentralized multi-modal learning system as shown in Fig. \ref{system}, which consists of a set of wireless devices $\mathcal{K}=\{1,2,...,K\}$. The set of modalities of device $k$ is $\mathcal{M}_k\subseteq \mathcal{M}$, where $\mathcal{M}=\{1,2,...,M\}$ contains all modalities in the considered system. Note that $\mathcal{M}_k$ varies across different devices. As shown in Fig. \ref{system}, smartphones possess data with two modalities, namely images and audio, whereas smart glasses exclusively comprise the image modality. Each device $k\in \mathcal{K}$ has its local dataset $\{(\{\bm{x}^m_{k,1}\}_{m\in\mathcal{M}_k},y_{k,1}),...,(\{\bm{x}^m_{k,D_k}\}_{m\in\mathcal{M}_k},y_{k,D_k})\}$ with the size $D_k$. $\bm{x}^m_{k,1},...,\bm{x}^m_{k,D_k}$ are the raw data of the modality $m\in\mathcal{M}_k$. $y_{k,1},...,y_{k,D_k}$ refer to the corresponding label. It is assumed that the data of different modalities on each device is temporally aligned.

The topology of the considered system can be regarded as an undirected graph. The set of neighboring devices of device $k$ is denoted as $\mathcal{R}_k$. Let $\mathcal{R}^m_k \subseteq \mathcal{R}_k$ be the set of neighboring devices that possess data with modality $m$. There exists a device-to-device (D2D) link between device $k$ and its neighboring device $k^{'}\in \mathcal{R}_k$, which can be used to transmit model parameters. After completing the local update in each training round, each device transmits the local model to its neighboring devices using frequency-division multiplexing. 

Through the collaborative learning of devices, the objective of this system is to find the optimal parameters $\bm{w}^{\ast}$ that minimizes the global loss function
\begin{equation}
	F(\bm{w})=\frac{1}{K}\sum_{k\in \mathcal{K}} F_k(\bm{w}).
\end{equation}
The component $F_k(\bm{w})=\frac{1}{D_k}\sum_{d=1}^{D_k}f(\{\bm{x}^m_{k,d}\}_{m\in\mathcal{M}_k},y_{k,d};\bm{w})$ is the task-specific loss function of device $k$, where $f(\cdot)$ is determined by the specific task, e.g., cross-entropy.

\begin{table}
	\centering
	\caption{Summary of Notations}
	\begin{tabular}{lp{6cm}}
		\hline
		Notation & Definition \\
		\hline
		$\mathcal{K}$ & Set of devices\\
		$\mathcal{M}_k$ & Set of modalities on device $k$\\
		$\bm{x}_{k,d}^m$, $y_{k,d}$ & Raw data for modality $m$ and label of $d$-th data on device $k$ \\
		$\mathcal{R}_k$ & Set of neighboring devices of device $k$ \\
		$\mathcal{R}^m_k$ & Set of neighboring devices that possess data with modality $m$ of device $k$ \\
		$\hat{\bm{h}}^{m}_{k,d}$, $\check{\bm{h}}^{m}_{k,d}$ & Modality-common and modality-specific feature extracted from the data of modality $m$\\
		$\hat{\bm{w}}^{n}_{k,t}$ & Modality-common classifier obtained by device $k$ after $n$-th iteration of $t$-th global round\\
		$\check{\bm{w}}^{n,m}_{k,t}$ & Modality-specific classifier of modality $m$ obtained by device $k$ after $n$-th iteration of $t$-th global round\\
		$\bm{w}^{n,m}_{k,t}$ & Feature extractor of modality $m$ obtained by device $k$ after $n$-th iteration of $t$-th global round\\
		$N_k$ & Number of local iterations in each round on device $k$\\
		$\bm{W}^m_{k,t}$,$\check{\bm{W}}^m_{k,t}$,$\hat{\bm{W}}_{k,t}$ & Aggregated parameter of $\bm{w}^{N_k,m}_{k,t}$, $\check{\bm{w}}^{N_k,m}_{k,t}$ and $\hat{\bm{w}}^{N_k}_{k,t}$ in $t$-th round on device $k$\\
		\hline
	\end{tabular}
	\label{sim_par}
\end{table}

\subsection{Multi-Modal Neural Networks with Feature Decomposition}
%For simplicity, we assume that all features $\bm{h}^m$ obtained from different modalities have the same length $L$.
In this paper, we take multi-modal neural networks with the widely used fusion method, i.e., concatenation \cite{OGM}. Fig. \ref{model} depicts an original multi-modal neural network that uses the concatenation fusion method. The data $\bm{x}^m$ of modality $m$ is processed using the network with the parameters $\bm{w}^m$ to extract the feature representation $\bm{h}^m$. The features from different modalities are concatenated, and then processed using the linear classifier with the weight $\tilde{\bm{w}}$ and the bias $\bm{b}$ to obtain the output $\bm{c}$, i.e.,
\begin{align}
	\bm{c}&=\tilde{\bm{w}} [(\bm{h}^1)^T,(\bm{h}^2)^T,...,(\bm{h}^M)^T]^T+\bm{b} \nonumber\\
	&=\sum_{m=1}^M\tilde{\bm{w}}^m \bm{h}^m +\bm{b}.
\end{align}
The classifier $\tilde{\bm{w}}\in \mathbb{R}^{C\times ML}$ can be decomposed into $M$ components corresponding to different modalities, i.e., $\tilde{\bm{w}}=[\tilde{\bm{w}}^1,\tilde{\bm{w}}^2,...,\tilde{\bm{w}}^M]$, where $\tilde{\bm{w}}^m \in \mathbb{R}^{C\times L}$. $C$ represents the number of categories in the dataset. $L$ is the length of the features $\bm{h}^m$, i.e., $\bm{h}^m \in \mathbb{R}^{L\times 1}$. The network in Fig. \ref{model} extracts a specific feature from each modality's data, which is not suitable for scenarios with modality absence. This is because a device lacking modality $m$ can not learn the correlated features between its own modalities and modality $m$. 

\begin{figure}[t]
	\centering
	\includegraphics[width=8cm]{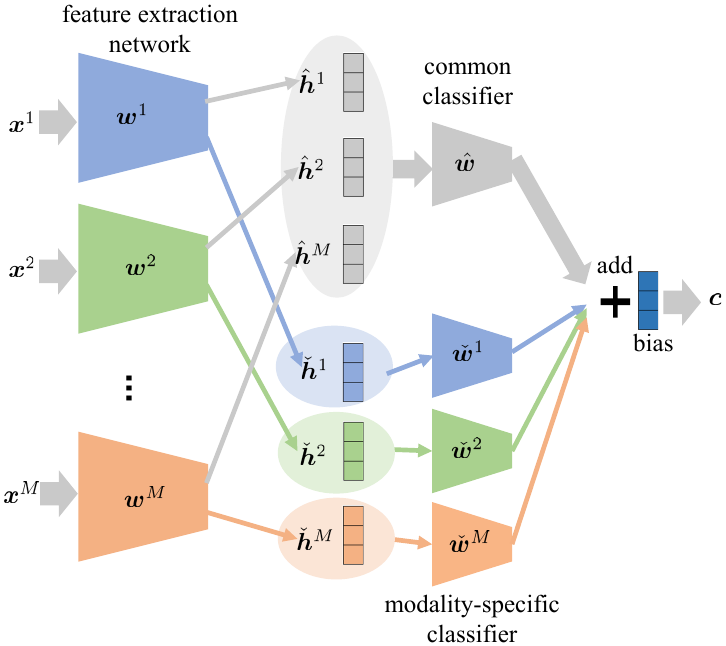}
	\caption{A multi-modal neural network with feature decomposition.}
	\label{model_KD}
\end{figure}

Therefore, as shown in Fig. \ref{model_KD}, we decompose the feature extracted from each modality into the modality-common component $\hat{\bm{h}}^{m}$ and the modality-specific component $\check{\bm{h}}^{m}$ based on the network in Fig. \ref{model}, i.e., 
\begin{align}
	[(\hat{\bm{h}}^{m})^T,(\check{\bm{h}}^{m})^T]^T=\phi(\bm{x}^{m};
	\bm{w}^{m}),
\end{align}
where $\phi(\cdot;\bm{w}^{m})$ refers to the feature extraction network with the parameter $\bm{w}^{m}$. The modality-common features extracted from the data of different modalities within the same category are expected to be in the same distribution. This facilitates the application of knowledge distillation to learn the correlated features between modalities in scenarios with modality absence. Subsequently, $\hat{\bm{h}}^{m}\in\mathbb{R}^{\hat{L}\times 1}$ and $\check{\bm{h}}^{m}\in\mathbb{R}^{\hat{L}\times 1}$ are processed by the common classifier $\hat{\bm{w}}$ and the modality-specific classifier $\check{\bm{w}}^{m}$ respectively, i.e.,
\begin{align}
	\label{decom}
	\bm{c}=\sum_{m=1}^M(\hat{\bm{w}}\hat{\bm{h}}^{m}+\check{\bm{w}}^{m}\check{\bm{h}}^{m})+\bm{b}.
\end{align}
The whole model is $\bm{w}=\{\bm{w}^1,\bm{w}^2,...,\bm{w}^M,\hat{\bm{w}}, \check{\bm{w}}^1,\check{\bm{w}}^2,...,\\ \check{\bm{w}}^M,\bm{b}\}$.

Due to the modality absence, only a subset of the model $\bm{w}$ is trained on each device. Specifically, device $k$ utilizes the data $\bm{x}_{k,d}^m$ from its available modalities $m\in \mathcal{M}_k$ to obtain the output $\bm{c}_{k,d}$, i.e.,
\begin{align}
	\bm{c}_{k,d}=\sum_{m\in \mathcal{M}_k}(\hat{\bm{w}}_{k,t}^n\hat{\bm{h}}_{k,d}^{m}+\check{\bm{w}}_{k,t}^{n,m}\check{\bm{h}}_{k,d}^{m}) +\bm{b}_{k,t}^{n},
\end{align}
where $\bm{w}^{n,m}_{k,t}$ and $\check{\bm{w}}_{k,t}^{n,m}$ are the parameter of the feature extraction network and the modality-specific classifier corresponding to modality $m$, respectively. $\hat{\bm{w}}_{k,t}^n$ is the parameter of the common classifier. $\bm{b}_{k,t}^{n}$ is the bias. They are obtained by device $k$ after the $n$-th iteration of the $t$-th global round. $[(\hat{\bm{h}}_{k,d}^{m})^T,(\check{\bm{h}}_{k,d}^{m})^T]^T$ is the feature extracted from $\bm{x}_{k,d}^m$ using the network with the parameter $\bm{w}^{n,m}_{k,t}$. To effectively decompose the modality-common feature $\hat{\bm{h}}_{k,d}^{m}$ and the modality-specific feature $\check{\bm{h}}_{k,d}^{m}$, the following loss functions are employed.

\subsubsection{Similarity Loss}
To make the modality-common features $\hat{\bm{h}}_{k,d}^{m}$ of different modalities be in the same distribution, we minimize the following loss function to approximate the distributions obtained by processing $\hat{\bm{h}}^{m}_{k,d}$ from different modalities using the common classifier.
\begin{align}
	\label{sim_loss}
	F^{sim}_k {\setlength\arraycolsep{0.5pt}=}\frac{1}{D_k}\sum_{d=1}^{D_k}\sum_{(m,m^{'})\in\hat{\mathcal{M}}_k}\frac{KL(\sigma(\hat{\bm{w}}^{n}_{k,t}\hat{\bm{h}}^{m}_{k,d})||\sigma(\hat{\bm{w}}^{n}_{k,t}\hat{\bm{h}}^{m^{'}}_{k,d}))}{|\mathcal{M}_k|(|\mathcal{M}_k|-1)},
\end{align}
where $\hat{\mathcal{M}}_k$ is the set of permutations formed by selecting any two different elements from $\mathcal{M}_k$, which has $|\mathcal{M}_k|(|\mathcal{M}_k|-1)$ elements. The number of elements in $\mathcal{M}_k$ is denoted by $|\mathcal{M}_k|$. $\sigma(\cdot)$ refers to the softmax function. KL-divergence \cite{KLD} can be used to measure the divergence between two different distributions $\bm{z}$ and $\hat{\bm{z}}$. It is defined by 
\begin{align}
	KL(\bm{z}||\hat{\bm{z}})=\sum_{i} z^i \ln \frac{z^i}{\hat{z}^i},
\end{align}
where $z^i$ and $\hat{z}^i$ represent the $i$-th component of $\bm{z}$ and $\hat{\bm{z}}$ respectively. The loss $F^{sim}_k$ can only be computed by devices that have multiple modalities. The devices with only one modality need the generator to guide their modality-common features towards the common distribution, which will be introduced in the following section.

\subsubsection{Auxiliary Classification Loss}
The distributions of the modality-common features for different data categories are distinct. To capture the category information within these features, we minimize the auxiliary classification loss as following
\begin{align}
	\label{cls_loss}
	F^{cls}_k=\frac{1}{D_k|\mathcal{M}_k|}\sum_{d=1}^{D_k}\sum_{m\in\mathcal{M}_k}CE(\sigma(\hat{\bm{w}}^{n}_{k,t}\hat{\bm{h}}^{m}_{k,d}),y_{k,d}),
\end{align}
where $CE(\sigma(\hat{\bm{w}}^{n}_{k,t}\hat{\bm{h}}^{m}_{k,d}),y_{k,d})$ is the cross-entropy loss calculated with $\sigma(\hat{\bm{w}}^{n}_{k,t}\hat{\bm{h}}^{m}_{k,d})$ and the label $y_{k,d}$. 

\subsubsection{Difference Loss}
To ensure that the modality-common and modality-specific features capture distinct aspects of the input data, we utilize the orthogonality loss as follows \cite{diff_loss}
\begin{align}
	\label{dif_loss}
	F^{dif}_k=\sum_{m\in\mathcal{M}_k}\|(\hat{\bm{H}}^{m}_{k})^T\check{\bm{H}}^{m}_{k}\|_2^2,
\end{align}
where $\hat{\bm{H}}^{m}_{k}{\setlength\arraycolsep{0.5pt}=}[\hat{\bm{h}}^{m}_{k,1},\hat{\bm{h}}^{m}_{k,2},...,\hat{\bm{h}}^{m}_{k,D_k}] \in \mathbb{R}^{\hat{L}\times D_k}$ and $\check{\bm{H}}^{m}_{k}=[\check{\bm{h}}^{m}_{k,1},\check{\bm{h}}^{m}_{k,2},...,\check{\bm{h}}^{m}_{k,D_k}]\in \mathbb{R}^{\hat{L}\times D_k}$.

Therefore, the loss function of the feature decomposition is summarized as follows
\begin{align}
	\label{dec_loss}
	F^{dec}_k=\alpha_1 F^{sim}_k+\alpha_2 F^{cls}_k+\alpha_3 F^{dif}_k,
\end{align}
where $\alpha_1$, $\alpha_2$ and $\alpha_3$ are hyper-parameters. This loss is minimized to optimize the parameters of the feature extractors and the common classifier, i.e., $\{\bm{w}^{n,m}_{k,t}\}_{m\in\mathcal{M}_k}\cup\{\hat{\bm{w}}^{n}_{k,t}\}$.

\begin{figure*}[t]
	\centering
	\includegraphics[width=15cm]{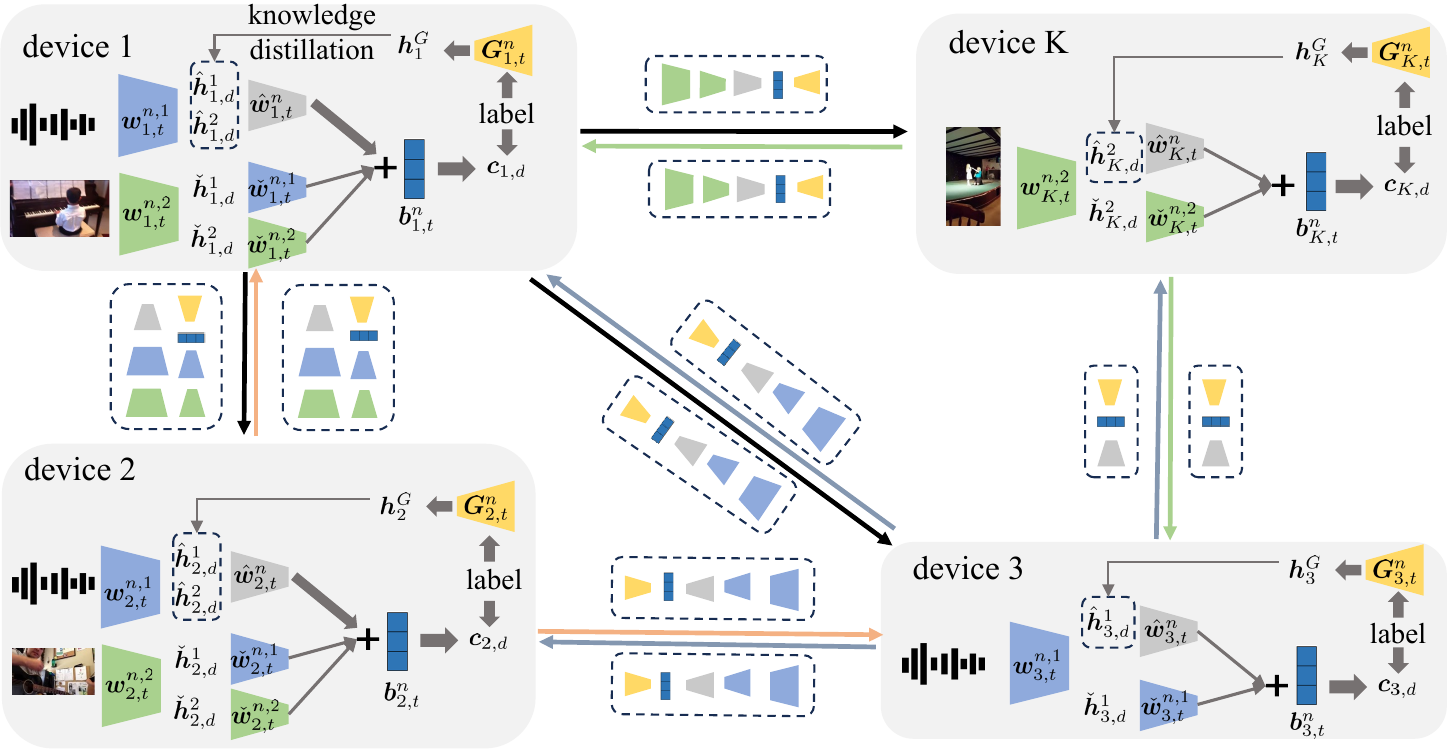}
	\caption{Proposed decentralized multi-modal learning with knowledge distillation.}
	\label{sys_KD}
\end{figure*}
\subsection{Decentralized Learning}
The execution process of decentralized learning can be divided into two stages. In the local update stage, each device updates the model parameter using its data. In the parameter aggregation stage, each device transmits the locally updated parameter to its neighboring devices and aggregates its parameter with the received parameters. The decentralized learning process for the multi-modal network presented in the above section is introduced as follows.

\subsubsection{Local Update Stage}
Due to the different modalities across devices, the model trained on different devices differ during the local update stage. Specifically, the model parameters $\bm{w}^{n,m}_{k,t},\check{\bm{w}}^{n,m}_{k,t}$ associated with the modality $m \in \mathcal{M}_k$ and the common parameters $\hat{\bm{w}}^{n}_{k,t},\bm{b}_{k,t}^{n}$ are used to calculate the output $\bm{c}_{k,d}$. These parameters are updated by device $k$. The parameters corresponding to the other modalities, i.e., $\{\bm{w}^{n,m}_{k,t},\check{\bm{w}}^{n,m}_{k,t}\}_{m \notin \mathcal{M}_k}$, are not used in the forward propagation (FP) and the backward propagation (BP), thus they are not updated. The parameter that can be updated by device $k$ is denoted by $\bm{w}^{n}_{k,t}=\{\bm{w}^{n,m}_{k,t},\check{\bm{w}}^{n,m}_{k,t}\}_{m\in \mathcal{M}_k}\cup\{\hat{\bm{w}}^{n}_{k,t},\bm{b}_{k,t}^{n}\}$. The update equation of the parameters is
\begin{align}
	\label{grad}
	\bm{w}^n_{k,t} =\bm{w}^{n-1}_{k,t}-\eta_t \nabla F^{tot}_k (\bm{w}^{n-1}_{k,t}),
\end{align}
for $n=1,2,...,N_k$. $N_k$ is the number of iterations of the local update, which is determined by the number of local epochs, the amount of local data and the batch size of device $k$. The total loss $F^{tot}_k$ about the training of the multi-modal neural network is given by
\begin{align}
	\label{tot_loss}
	F^{tot}_k=F_k+F^{dec}_k+\alpha_4 F^{kd}_k,
\end{align}
where $\alpha_4$ is a hyper-parameter. The loss $F^{kd}_k$ about knowledge distillation is introduced in the subsequent section. The initial parameter of device $k$ in the $t$-th global round is $\bm{w}^0_{k,t}=\bm{W}_{k,t-1}$, where $\bm{W}_{k,t-1}$ is the aggregated model parameter of device $k$ obtained in the $(t-1)$-th round.

\subsubsection{Parameter Aggregation Stage}
In the parameter aggregation stage, device $k$ sends the parameters $\bm{w}^{N_k,m}_{k,t}$ and $\check{\bm{w}}^{N_k,m}_{k,t}$ to its neighboring devices that possess data with modality $m$, i.e., $k^{'} \in \mathcal{R}_k^m$. In contrast, the parameters $\hat{\bm{w}}^{N_k}_{k,t}, \bm{b}_{k,t}^{N_k}$ are transmitted to all the neighboring devices, i.e., $k^{'} \in \mathcal{R}_k$. The parameters specific to modality $m$ can only be exchanged between neighboring devices that possess data with modality $m$ because the feature extraction networks for different modalities are distinct. Then device $k$ aggregates its parameters with the received parameters by
\begin{align}
	\label{avg}
	&\bm{W}^m_{k,t}{\setlength\arraycolsep{0.5pt}=}\sum_{k^{'}\in \mathcal{R}_k^m \cup \{k\}} \xi^m_{k,k^{'}} \bm{w}^{N_{k^{'}},m}_{k^{'},t},\nonumber\\
	&\check{\bm{W}}^m_{k,t}{\setlength\arraycolsep{0.5pt}=}\sum_{k^{'} \in \mathcal{R}_k^m \cup \{k\}} \xi^m_{k,k^{'}} \check{\bm{w}}^{N_{k^{'}},m}_{k^{'},t}, \nonumber\\
	&\hat{\bm{W}}_{k,t}=\sum_{k^{'} \in \mathcal{R}_k \cup \{k\}} \xi_{k,k^{'}} \hat{\bm{w}}^{N_{k^{'}}}_{k^{'},t}, \nonumber\\
	&\bm{B}_{k,t}=\sum_{k^{'} \in \mathcal{R}_k \cup \{k\}} \xi_{k,k^{'}} \bm{b}^{N_{k^{'}}}_{k^{'},t},
\end{align}
where $\xi^m_{k,k^{'}}\geq0$ and $\xi_{k,k^{'}}\geq0$ are the aggregation coefficients for the parameters specific to modality $m$ and the common parameters, respectively. They satisfy $\sum_{k=1}^K\xi^m_{k,k^{'}}=1$, $\sum_{k^{'}=1}^K\xi^m_{k,k^{'}}=1$, $\sum_{k=1}^K\xi_{k,k^{'}}=1$ and $\sum_{k^{'}=1}^K\xi_{k,k^{'}}=1$. In particular, $\xi^m_{k,k^{'}}=0$ for $k^{'} \notin \mathcal{R}_k^m \cup \{k\}$, and $\xi_{k,k^{'}}=0$ for $k^{'} \notin \mathcal{R}_k \cup \{k\}$. The aggregated parameter $\bm{W}_{k,t}=\{\bm{W}^m_{k,t},\check{\bm{W}}^m_{k,t}\}_{m\in \mathcal{M}_k}\cup\{\hat{\bm{W}}_{k,t}, \bm{B}_{k,t}\}$ is used as the initial parameter for the next local update stage of device $k$. The local update and parameter aggregation are performed for many global rounds to obtain the desired learning performance.

If decentralized training is conducted solely using the task-specific loss function $F_k$, device $k$ can only learn the correlated features among its own modalities $m\in \mathcal{M}_k$ and cannot acquire the features related to the other modalities $m\notin \mathcal{M}_k$, which degrades the performance of DMML. To address this issue, we apply the knowledge distillation to DMML.

\subsection{DMML with Knowledge Distillation}
While some works have integrated knowledge distillation with federated learning \cite{fed_KD,data_free}, to the best of our knowledge, knowledge distillation has not yet been applied to DMML.  

As shown in Fig. \ref{sys_KD}, in the proposed training scheme, each device can utilize a generator to learn the conditional distribution of the modality-common features under different categories. Due to the non-IID data and modality heterogeneity, the locally-trained generator of each device can only learn the local distribution. By means of parameter exchange and aggregation, the generator can learn the global conditional distribution of the modality-common features. Consequently, devices can utilize the generator to guide their modality-common features towards the global distribution. In other words, a part of features from each modality can be guided towards a common distribution, thereby potentially capturing correlated features across different modalities.

\subsubsection{Knowledge Extraction}
To make the generator $\bm{G}^{n}_{k,t}$ learn the knowledge about the distribution of the modality-common features, each device utilizes the trained common classifier $\hat{\bm{w}}^{N_k}_{k,t}$ to guide the training of the generator with the following loss
\begin{align}
	\label{gen_loss}
	F^{gen}_k= \mathbb{E}_{\substack {\bm{h}^{G}_k=\bm{G}^{n}_{k,t}(\bm{z},y),\\ \bm{z}\sim\mathcal{N}(0,\bm{I}),y\in \mathcal{C}_k}}[CE(\sigma(\hat{\bm{w}}^{N_k}_{k,t}\bm{h}^{G}_k),y) {\setlength\arraycolsep{0.5pt}+} \beta \|\bm{h}^{G}_k-\bar{\bm{h}}^y_{k,t} \|_2],
\end{align}
where $\mathcal{C}_k$ is the set of labels possessed by device $k$. Given a label $y\in \mathcal{C}_k$, the generator $\bm{G}^{n}_{k,t}$ can map a Gaussian random variable $\bm{z}\sim\mathcal{N}(0,\bm{I})$ to a feature $\bm{h}^{G}_k$. $\beta$ is a hyper-parameter. By minimizing the loss function $F^{gen}_k$, the generated feature $\bm{h}^{G}_k$ is expected to be correctly classified by the common classifier, while approaching the mean of the true features, i.e., $\bar{\bm{h}}^y_{k,t}=\frac{1}{|Y_k^y||\mathcal{M}_k|}\sum_{d\in Y_k^y}\sum_{m\in\mathcal{M}_k}\hat{\bm{h}}^m_{k,d}$. The set $Y_k^y=\{d|y_{k,d}=y\}$ contains the indices of data with the label $y$. This enables the generator to learn the conditional distribution of the modality-common features. 

The generator locally trained on each device can only learn the local distribution because of the non-IID data and modality heterogeneity. To learn the global conditional distribution of the modality-common features, each device can send its generator $\bm{G}^{\hat{N}_k}_{k,t}$ to its neighboring devices and aggregate its generator with the received generators.
\begin{align}
	\label{kd_avg}
	\bar{\bm{G}}_{k,t}=\sum_{k^{'} \in \mathcal{R}_k\cup \{k\}} \xi_{k,k^{'}} \bm{G}^{\hat{N}_k}_{k^{'},t},
\end{align}
where $\hat{N}_k$ is the number of local training iterations for the generator of device $k$. 

\subsubsection{Knowledge Distillation}
In the next training round, the devices can utilize the aggregated generator $\bar{\bm{G}}_{k,t}$ to guide its modality-common features towards the global distribution by minimizing the following loss
\begin{align}
	\label{kd_loss}
	F^{kd}_k{\setlength\arraycolsep{0.5pt}=} \frac{1}{D_k|\mathcal{M}_k|}\sum_{d=1}^{D_k}\sum_{m\in\mathcal{M}_k}&\mathbb{E}_{\substack {\bm{h}^{G}_k=\bar{\bm{G}}_{k,t}(\bm{z},y_{k,d}),\\ \bm{z}\sim\mathcal{N}(0,\bm{I})}}[KL(\sigma(\hat{\bm{w}}^{n}_{k,t+1}\hat{\bm{h}}^m_{k,d})\nonumber \\
	&\qquad\qquad ||\sigma(\hat{\bm{w}}^{n}_{k,t+1}\bm{h}^{G}_k))].
\end{align}
Minimizing the divergence between $\sigma(\hat{\bm{w}}^{n}_{k,t+1}\bm{h}_{k,d})$ and $\sigma(\hat{\bm{w}}^{n}_{k,t+1}\bm{h}^{G}_k)$ enables the modality-common features of device $k$ to approach the common distribution learned by the generator. The overall training process of the proposed DMML-KD is outlined in Algorithm \ref{alg_outline}. 

\begin{algorithm}[t]
	\caption{Proposed decentralized multi-modal learning with knowledge distillation}
	\label{alg_outline}
	\hspace*{0.02in}{\bf Input:}
	$\bullet$ The initial parameters of the multi-modal network $\bm{W}_{k,0}$ and the generator $\bar{\bm{G}}_{k,0}$ for all devices\\
	$\bullet$ The learning rate $\eta_t$ and $\eta_t^G$ of the multi-modal network and the generator\\
	$\bullet$ The aggregation coefficients $\xi^m_{k,k^{'}}$ and $\xi_{k,k^{'}}$ for all devices
	
	\begin{algorithmic}[1]
		\FOR {training round $t=1$ to $T$}
		\FOR {all devices $k=1$ to $K$ in parallel}
		\STATE Set $\bm{w}^0_{k,t}=\bm{W}_{k,t-1}$, $\bm{G}^0_{k,t}=\bar{\bm{G}}_{k,t-1}$
		\FOR {iteartion $n=1$ to $N_k$}
		\STATE Train the multi-modal network with $F^{tot}_k$
		\ENDFOR
		\STATE	Calculate the average feature $\bar{\bm{h}}_{k,t}^y$
		\FOR {iteartion $n=1$ to $\hat{N}_k$}
		\STATE Use $\hat{\bm{w}}^{N_k}_{k,t}$ and $\bar{\bm{h}}_{k,t}^y$ to train the generator with $F^{gen}_k$ 
		\ENDFOR
		\STATE Transmit the parameters $\bm{w}^{N_k,m}_{k,t}$ and $\check{\bm{w}}^{N_k,m}_{k,t}$ to its neighboring devices $k^{'} \in \mathcal{R}_k^m$
		\STATE Transmit the parameters $\hat{\bm{w}}^{N_k}_{k,t}, \bm{b}_{k,t}^{N_k}, \bm{G}_{k,t}^{\hat{N}_k}$ to all the neighboring devices $k^{'} \in \mathcal{R}_k$
		\STATE Aggregate the parameters with (\ref{avg}) and (\ref{kd_avg})
		\ENDFOR
		\ENDFOR
	\end{algorithmic}
\end{algorithm}

%%%%%%%%%%%%%%%%%%%%%%%%%%%%%%%%%%%%%%%
\section{Training Balance and Problem Formulation}
\label{prob}
To address the issue of imbalanced training across different modalities, we propose a method that controls the number of local iterations for different modalities. Compared to adjusting the learning rates \cite{OGM,balan_LR} of different modalities, controlling the number of iterations can reduce the computational load of devices, thereby reducing the energy consumption and the training latency. To achieve the balanced training by controlling the number of iterations, we introduce a novel performance metric based on the parameter variation, which can evaluate the training speed of different modalities. Furthermore, we formulate an optimization problem to optimize the number of iterations for different modalities in each training round.

\subsection{Control over the Number of Iterations}
In the local update process introduced in Section \ref{sys_model}, device $k$ performs joint training in each iteration using the data from all the modalities it possesses. In other words, the parameters $\bm{w}^{n,m}_{k,t},\check{\bm{w}}^{n,m}_{k,t}$ corresponding to the modalities in $\mathcal{M}_k$ are updated in each iteration. Due to the different training speeds of different modalities, it is unreasonable to impose the same number of iterations on different modalities. In this paper, we reduce the number of iterations for the modalities that exhibit faster training speed, thereby achieving the training balance across different modalities.

Specifically, we use $P_{k,t}^{n,m}=1$ to denote that the parameters corresponding to the modality $m$ are trained by the device $k$ in the $n$-th iteration of the $t$-th round, otherwise $P_{k,t}^{n,m}=0$. Therefore, the output of device $k$ in the $n$-th iteration of the $t$-th round is given by
\begin{align}
	\label{control_it}
	\bm{c}_{k,d}=\sum_{m\in \mathcal{M}_k}P_{k,t}^{n,m}(\hat{\bm{w}}_{k,t}^{n-1}\hat{\bm{h}}_{k,d}^{m}+\check{\bm{w}}_{k,t}^{n-1,m}\check{\bm{h}}_{k,d}^{m}) +\bm{b}_{k,t}^{n-1}.
\end{align}
If $P_{k,t}^{n,m}=0$, the feature extraction network $\bm{w}_{k,t}^{n-1,m}$ corresponding to modality $m$ for device $k$ is not utilized to extract features in the $n$-th iteration. Because the features $[(\hat{\bm{h}}_{k,d}^{m})^T,(\check{\bm{h}}_{k,d}^{m})^T]^T$ is not extracted, the classifiers $\hat{\bm{w}}_{k,t}^{n-1}, \check{\bm{w}}_{k,t}^{n-1,m}$ do not process them. Therefore, the parameters $\bm{w}_{k,t}^{n-1,m}, \check{\bm{w}}_{k,t}^{n-1,m}$ corresponding to modality $m$ is not updated, i.e., $\bm{w}_{k,t}^{n,m}=\bm{w}_{k,t}^{n-1,m}$ and $\check{\bm{w}}_{k,t}^{n,m}=\check{\bm{w}}_{k,t}^{n-1,m}$. The common classifier $\hat{\bm{w}}_{k,t}^{n-1}$ can be updated because it may process the modality-common features from other modalities. If $P_{k,t}^{n,m}=0$ for all the modalities in $\mathcal{M}_k$, the whole model $\{\bm{w}^{n-1b,m}_{k,t},\check{\bm{w}}^{n-1,m}_{k,t}\}_{m\in \mathcal{M}_k}\cup\{\hat{\bm{w}}^{n-1}_{k,t},\bm{b}_{k,t}^{n-1}\}$ is not updated in the $n$-th iteration. 

We use the integer $0\leq N^m_{k,t}\leq N_{k}$ to represent the number of iterations performed by the device $k$ on the parameters corresponding to the modality $m$ in the $t$-th round. $N^m_{k,t}$ is obtained with the metric and optimization problem described later. Once $N^m_{k,t}$ is obtained, the value of $P_{k,t}^{n,m}$ for each iteration is randomly selected from $\{0, 1\}$ under the condition that $\sum_{n=1}^{N_k}P_{k,t}^{n,m}=N^m_{k,t}$.

Different from the parameters $\{\bm{w}^{n,m}_{k,t},\check{\bm{w}}^{n,m}_{k,t}\}_{m\notin \mathcal{M}_k}$, which remain unchanged throughout the local update of device $k$, the parameters $\{\bm{w}^{n,m}_{k,t},\check{\bm{w}}^{n,m}_{k,t}\}_{m\in \mathcal{M}_k}$ are not updated in a part of local iterations. The updated parameters $\{\bm{w}^{N_k,m}_{k,t},\check{\bm{w}}^{N_k,m}_{k,t}\}_{m\in \mathcal{M}_k}$ need to be transmitted to the neighboring devices in $\mathcal{R}_k^m$. 

%In particular, if $N^m_{k,t}=0$, the parameters $\bm{w}^{n,m}_{k,t},\check{\bm{w}}^{n,m}_{k,t}$ are not updated in the $t$-th round, and $\bm{w}^{N_k,m}_{k,t},\check{\bm{w}}^{N_k,m}_{k,t}$ are not required to be transmitted. 

\subsection{Performance Metric for Training Balance}
In this paper, we evaluate the training speed of different modalities by measuring the relative parameter variation of the feature extractor for each modality, i.e.,
\begin{align}
	\label{par_var}
	&\frac{\|\bm{W}_{k,t}^m-\bm{W}_{k,t-1}^m\|_2}{\|\bm{W}_{k,t}^m\|_2}\nonumber\\
	&=\frac{\|\sum_{k^{'} =1}^K \xi^m_{k,k^{'}} \bm{w}^{N_{k^{'}},m}_{k^{'},t}-\bm{W}_{k,t-1}^m\|_2}{\|\bm{W}_{k,t}^m\|_2}\nonumber\\
	& \stackrel{(a)}{\leq} \sum_{k^{'} =1}^K \xi^m_{k,k^{'}}\frac{\| \bm{w}^{N_{k^{'}},m}_{k^{'},t}-\bm{W}_{k,t-1}^m\|_2}{\|\bm{W}_{k,t}^m\|_2}
\end{align}
where step (a) is due to the convexity of $\|\cdot\|_2$ and $\sum_{k^{'} =1}^K \xi^m_{k,k^{'}}=1$. We use $\varphi_{k,k^{'},t}^m=\frac{\| \bm{w}^{N_{k^{'}},m}_{k^{'},t}-\bm{W}_{k,t-1}^m\|_2}{\|\bm{W}_{k,t}^m\|_2}$ to denote the relative parameter variation between the parameter $\bm{w}^{N_{k^{'}},m}_{k^{'},t}$ obtained by the locally update of device $k^{'}$ and the aggregated parameter $\bm{W}_{k,t-1}^m$ of device $k$. 

Generally, a larger loss function leads to larger gradients, resulting in bigger parameter variations. As training approaches convergence, the parameters tend to stabilize, indicating smaller variations in parameters. If the parameter variation of the parameters corresponding to modality $m$ is smaller than that of the other modalities, the loss function for modality $m$ is smaller and closer to convergence. Hence, the number of iterations corresponding to modality $m$ needs to be reduced. This implies that the number of iterations should be inversely proportional to the magnitude of parameter variations.

However, we cannot directly determine the number of iterations for modality $m$ on device $k$ based on the parameter variation $\frac{\|\bm{W}_{k,t}^m-\bm{W}_{k,t-1}^m\|_2}{\|\bm{W}_{k,t}^m\|_2}$. This is because the parameter variation $\frac{\|\bm{W}_{k,t}^m-\bm{W}_{k,t-1}^m\|_2}{\|\bm{W}_{k,t}^m\|_2}$ of device $k$ is influenced not only by its local update, i.e., $\varphi_{k,k,t}^m$, but also by the parameter update of its neighboring devices, i.e., $\varphi_{k,k^{'},t}^m, k^{'}\in\mathcal{R}_k^m$, as shown in (\ref{par_var}). These influences are combined with the aggregation coefficient $\xi^m_{k,k^{'}}$. For modality $m$ with the small parameter variation on device $k$, it is necessary to not only reduce the number of iterations for modality $m$ on device $k$, but also consider reducing the number of iterations for modality $m$ on the neighboring device $k^{'}\in\mathcal{R}_k^m$. Meanwhile, when reducing the number of iterations for modality $m$ on device $k$, it is important to consider the impact on other devices after the parameter aggregation, which can be quantified by the parameter variation  $\varphi_{k,k^{'},t}^m$.

Based on the above analysis, we develop the following performance metric of training balance
\begin{align}
	\label{bal_metric}
	\gamma_{t+1}^m&=\sum_{k=1}^K\sum_{k^{'} =1}^K \xi^m_{k,k^{'}} \frac{N^m_{k^{'},t+1}}{\varphi_{k,k^{'},t}^m+\varepsilon}\nonumber\\
	&=\sum_{k^{'} =1}^KN^m_{k^{'},t+1}\sum_{k=1}^K  \frac{\xi^m_{k,k^{'}}}{\varphi_{k,k^{'},t}^m+\varepsilon},
\end{align}
where $\varepsilon>0$ is a very small number to avoid division by zero. After the parameter aggregation in the $t$-th round, the number of iterations $N^m_{k^{'},t+1}$ of the $(t+1)$-th round is optimized to make $\gamma_{t+1}^m$ be close for different modalities. It can be observed from (\ref{bal_metric}) that if the parameter variation $\varphi_{k,k^{'},t}^m$ is big, the number of iterations is expected to be high. The term $N^m_{k^{'},t+1}\sum_{k=1}^K  \frac{\xi^m_{k,k^{'}}}{\varphi_{k,k^{'},t}^m+\varepsilon}$ can be regarded as the influence of device $k^{'}$, after $N^m_{k^{'},t+1}$ iterations, on other devices that require aggregation with $\bm{w}^{N_{k^{'}},m}_{k^{'},t+1}$. 

\subsection{Problem Formulation}
Optimizing the number of iterations $N^m_{k,t}$ can have an impact on the energy consumption of devices. The energy consumption of devices is primarily attributed to two factors: the local update and the parameter transmission. Specifically, the energy consumption of device $k$ in the $t$-th round is given by \cite{burd1996processor}
\begin{align}
	\label{tot_ene}
	&E_{k,t}=E^{cmp}_{k,t}+E^{com}_{k,t}\nonumber\\
	&=\frac{\epsilon_kf_k^2}{e_k}(\sum_{m\in\mathcal{M}_k} N^m_{k,t}(O^m{\setlength\arraycolsep{0.5pt}+}\hat{O}){\setlength\arraycolsep{0.5pt}+}N^{\max}_{k,t}(O^b{\setlength\arraycolsep{0.5pt}+}O^G_F){\setlength\arraycolsep{0.5pt}+}\hat{N}_{k}O^G)\nonumber\\
	&\qquad+\sum_{m\in\mathcal{M}_k}\sum_{k^{'} \in \mathcal{R}^m_k}\frac{p_k\Upsilon^m}{\mathcal{B}\log(1{\setlength\arraycolsep{0.5pt}+}\frac{p_kg_{k,k^{'},t}^2}{\mathcal{B}\mathcal{N}_0})} \nonumber\\
	&\qquad+\sum_{k^{'} \in \mathcal{R}_k}\frac{p_k\Upsilon}{\mathcal{B}\log(1{\setlength\arraycolsep{0.5pt}+}\frac{p_kg_{k,k^{'},t}^2}{\mathcal{B}\mathcal{N}_0})},
\end{align}
where $O^m$ is the total floating point operations (FLOPs) of performing the FP and BP for the feature extractor and classifier specific to modality $m$ in each iteration. $\hat{O}$ and $O^b$ represent the total FLOPs of the FP and BP for the common classifier and the bias in each iteration, respectively. The FLOPs of the FP for the generator are denoted as $O^G_F$ in each iteration, while the total FLOPs of FP and BP for the generator is denoted as $O^G$ in each iteration. $f_k$ and $e_k$ are the computing frequency and the number of FLOPs per cycle for device $k$, respectively. $\epsilon_k$ is the effective capacitance coefficient of computing chip for device $k$. $N^{\max}_{k,t}$ is the maximum number of iterations that device $k$ can perform under the constraints of $N_k$ and the remaining energy. $\Upsilon$ denotes the total size (\emph{in bit}) of the common classifier, the bias and the generator. $\Upsilon^m$ denotes the total size of the parameters specific to modality $m$. Device $k$ transmits the parameters to its neighbors using a channel with a bandwidth of $\mathcal{B}$ through the transmission power $p_k$. The channel gain between device $k$ and $k^{'}$ in the $t$-th round is $g_{k,k^{'},t}$. The spectral density of the additive white Gaussian noise (AWGN) is $\mathcal{N}_0$. 

%The indicator $\mathbbm{1}_{k,t}^m=1$ if $N^m_{k,t}\neq0$, otherwise, $\mathbbm{1}_{k,t}^m=0$. 

In this paper, we optimize the number of iterations for the $(t+1)$-th round to achieve balanced training across devices for different modalities. The optimization problem can be formulated as
\begin{align}
\mathcal{P}:\min_{\{N^m_{k,t+1}\}}~ & \frac{\sum_{m=1}^M(\gamma_{t+1}^m-\frac{1}{M}\sum_{m^{'}=1}^M\gamma_{t+1}^{m^{'}})^2}{\sum_{k=1}^{K}\sum_{m\in\mathcal{M}_k}N^m_{k,t+1}} \nonumber \\
	s.t. ~ & E_{k,t+1} \leq E_{k,t+1}^{\max}, \forall k, \label{energy}\\
	& 0 \leq N^m_{k,t+1} \leq N_k, N^m_{k,t+1} \in \mathbb{Z}, \forall m \in \mathcal{M}_k, \forall k \label{iter_var},\\
	& \sum_{k^{'}\in \mathcal{R}_k^m} (\Delta N^m_{k,t+1}{\setlength\arraycolsep{0.5pt}-}\Delta N^m_{k^{'},t+1})^2{\setlength\arraycolsep{0.5pt}\leq} \delta, \forall m \in \mathcal{M}_k, \forall k. \label{reduc_eq}
\end{align}
The constraint (\ref{energy}) represents that the energy consumption $E_{k,t+1}$ cannot exceed the remaining energy $E_{k,t+1}^{\max}=E_{k,t}^{\max}-E_{k,t}$. The constraint (\ref{iter_var}) denotes the feasible region of $N^m_{k,t+1}$. The term $\Delta N^m_{k,t+1}=N^{\max}_{k,t+1}-N^m_{k,t+1}$ denotes the number of iterations reduced for modality $m$ of device $k$ in the $(t+1)$-th round. A larger value of $\Delta N^m_{k,t+1}$ indicates greater energy saving of device $k$. For fair consideration, it is expected that the saved number of iterations of different devices are similar, as shown by the constraint (\ref{reduc_eq}). The optimization objective is to minimize the variances of $\gamma_{k,t+1}^m$ across all modalities. Meanwhile, the total number of iterations across all devices is maximized to avoid slow convergence due to insufficient iteration per round. After each round of parameter aggregation, it is required to solve this optimization problem to obtain the number of iterations $N^m_{k^{'},t+1}$ in the next round.

%%%%%%%%%%%%%%%%%%%%%%%%%%%%%%%%%%%%%%%
\section{Training Balance Algorithm Design}
\label{alg}
The optimization problem $\mathcal{P}$ is non-convex. Therefore, we develop a low-complexity algorithm to obtain the approximate solution $N^m_{k,t+1}$ in this section, and introduce how to obtain the solution in practical systems.
\subsection{Approximate Solution of Problem $\mathcal{P}$}
We first estimate the maximum number of iterations $N^{\max}_{k,t+1}$ that device $k$ can perform under the constraints (\ref{energy}) and (\ref{iter_var}). If device $k$ has sufficient energy, i.e., the energy consumption $E_{k,t+1}$ calculated using (\ref{tot_ene}) with $N_{k,t+1}^m=N_k$ is less than $E_{k,t+1}^{\max}$, then we have $N^{\max}_{k,t+1}=N_k$. Otherwise, $N^{\max}_{k,t+1}$ is the maximum number of iterations that can be performed with the device's remaining energy, i.e., $N^{\max}_{k,t+1}$ is the maximum integer that satisfies the condition where the energy consumption $E_{k,t+1}$ calculated with $N_{k,t+1}^m=N^{\max}_{k,t+1}$ is less than $E_{k,t+1}^{\max}$.

Then the variances of $\gamma_{k,t+1}^m$ across all modalities are taken into consideration, which can be minimized with the following condition
\begin{align}
	\label{eq_gamma}
	\gamma_{t+1}^m=\frac{1}{M}\sum_{m^{'}=1}^M\gamma_{t+1}^{m^{'}},\forall m.
\end{align}
The condition (\ref{eq_gamma}) indicates that $\gamma_{t+1}^m=\gamma_{t+1}^{m^{'}}$, $\forall m, m^{'}$. From (\ref{bal_metric}), we can observe that $\gamma_{t+1}^m$ is determined by the number of iterations corresponding to modality $m$, i.e., $N_{k,t+1}^m, \forall k$. By maximizing $\sum_{k=1}^{K}\sum_{m\in\mathcal{M}_k}N^m_{k,t+1}$ while satisfying the condition (\ref{eq_gamma}), we can obtain the modality that uses the maximum feasible number of iterations, i.e.,
\begin{align}
	\label{max_modal}
	\hat{m}=\arg\min_{m}\sum_{k^{'} =1}^K N^{\max}_{k^{'},t+1}\sum_{k=1}^K  \frac{\xi^m_{k,k^{'}}}{\varphi_{k,k^{'},t}^m+\varepsilon}.
\end{align}

The number of iterations corresponding to the other modalities $m\neq \hat{m}$ should be reduced in order to make $\gamma_{t+1}^m$ close to $\gamma_{t+1}^{\hat{m}}$. To satisfy the constraint (\ref{reduc_eq}), which is to make $\Delta N^m_{k,t+1}$ similar across different devices, we employ a method of uniformly decreasing the number of iterations. As shown in Algorithm \ref{obtain_iter}, the initial number of iterations for each device $k\in \mathcal{K}^m=\{k|m\in\mathcal{M}_k,|\mathcal{M}_k|>1\}$ is set as $N^{\max}_{k,t+1}$, then the iterations for the devices $k\in \mathcal{K}^m$ are successively decreased until the following condition holds 
\begin{align}
	\label{end_con}
	\sum_{k^{'} =1}^K N^m_{k^{'},t+1}\sum_{k=1}^K  \frac{\xi^m_{k,k^{'}}}{\varphi_{k,k^{'},t}^m+\varepsilon}\leq \gamma_{t+1}^{\hat{m}}.
\end{align}
Since the devices with only one modality do not suffer from the training imbalance across modalities, their iteration count is set as $N^{\max}_{k,t+1}$.

\begin{algorithm}[t]
	\caption{Obtain the number of iterations}
	\label{obtain_iter}
	\hspace*{0.02in}{\bf Input:}
	The maximum number of iterations $N^{\max}_{k,t+1}, \forall k$; The coefficients $\frac{\xi^m_{k,k^{'}}}{\varphi_{k,k^{'},t}^m+\varepsilon},\forall m\in\mathcal{M}_k,\forall k, \forall k^{'}$.
	
	\begin{algorithmic}[1]
		\STATE	Obtain $\hat{m}$ with (\ref{max_modal})
		\FOR {$m=1$ to $M$}
		\STATE Set $N^m_{k,t+1}=N^{\max}_{k,t+1}, \forall k \in \mathcal{K}^m$
		\REPEAT
		\STATE Select the next device $k \in \mathcal{K}^m$
		\STATE Let $N^m_{k,t+1}-1\rightarrow N^m_{k,t+1}$
		\UNTIL{the condition (\ref{end_con}) holds}
		\ENDFOR
	\end{algorithmic}
	\hspace*{0.02in}{\bf Output:}
	The number of iterations $N^{m}_{k,t+1}, \forall m \in\mathcal{M}_k,\forall k$
\end{algorithm}
\begin{algorithm}[t]
	\caption{Proposed DMML-KD with training balance}
	\label{DMML-TB}
	\hspace*{0.02in}{\bf Input:}
	The initial parameters $\bm{W}_{k,0}$, $\bar{\bm{G}}_{k,0}$; The learning rate $\eta_t$ and $\eta_t^G$; The aggregation coefficients $\xi^m_{k,k^{'}}$ and $\xi_{k,k^{'}}$.
	
	\begin{algorithmic}[1]
		\FOR {training round $t=1$ to $T$}
		\FOR {all devices $k=1$ to $K$ in parallel}
		\STATE Set $\bm{w}^0_{k,t}=\bm{W}_{k,t-1}$, $\bm{G}^0_{k,t}=\bar{\bm{G}}_{k,t-1}$
		\STATE Randomly choose $P_{k,t}^{n,m}$ with $\sum_{n=1}^{N_k}P_{k,t}^{n,m}=N^m_{k,t}$
		\FOR {iteartion $n=1$ to $N_k$}
		\STATE Train the multi-modal networks with $P_{k,t}^{n,m}=1$ using $F^{tot}_k$
		\ENDFOR
		\STATE	Calculate the average feature $\bar{\bm{h}}_{k,t}^y$
		\FOR {iteartion $n=1$ to $\hat{N}_k$}
		\STATE Use $\hat{\bm{w}}^{N_k}_{k,t}$ and $\bar{\bm{h}}_{k,t}^y$ to train the generator with $F^{gen}_k$ 
		\ENDFOR
		\STATE Transmit $\bm{w}^{N_k,m}_{k,t}$ and $\check{\bm{w}}^{N_k,m}_{k,t}$ to devices $k^{'} \in \mathcal{R}_k^m$. Transmit $\hat{\bm{w}}^{N_k}_{k,t}, \bm{b}_{k,t}^{N_k}, \bm{G}_{k,t}^{\hat{N}_k}$ to devices $k^{'} \in \mathcal{R}_k$
		\STATE Aggregate the parameters with (\ref{avg}) and (\ref{kd_avg})
		\STATE Obtain the coefficients $N^{\max}_{k,t+1}$ and $\varphi_{k,k^{'},t}^m$. 
		\STATE Send $N^{\max}_{k,t+1}$ and $\frac{\xi^m_{k,k^{'}}}{\varphi_{k,k^{'},t}^m+\varepsilon},\forall m\in\mathcal{M}_k,\forall k^{'}$ to the cluster head
		\ENDFOR
		\STATE The cluster head utilizes Algorithm \ref{obtain_iter} to obtain $N^{m}_{k,t+1}$ and sends them back to the respective devices
		\ENDFOR
	\end{algorithmic}
\end{algorithm}

\subsection{Training Balance in Practice}
Based on the above description, we can observe that obtaining the iterations $N^{m}_{k,t+1}, \forall m \in\mathcal{M}_k,\forall k$ requires the coefficients $N^{\max}_{k,t+1}$ and $\frac{\xi^m_{k,k^{'}}}{\varphi_{k,k^{'},t}^m},\forall k, \forall k^{'},\forall m$. The maximum iteration $N^{\max}_{k,t+1}$ can be obtained by device $k$. The aggregation coefficients $\xi^m_{k,k^{'}}, \forall k^{'},\forall m$ are stored on device $k$. The parameter variation $\varphi_{k,k^{'},t}^m,\forall k^{'},\forall m$ can be calculated by device $k$ after receiving the parameters from other devices in the parameter aggregation stage of the $t$-th round. To obtain the number of iterations, each device needs to send its coefficients to the cluster head, which can be the device connected to the maximum number of devices. After receiving the coefficients sent by the other devices, the cluster head utilizes Algorithm \ref{obtain_iter} to obtain the number of iterations for different modalities on each device and sends them back to the respective devices. The overall process of the proposed DMML-KD with training balance is outlined in Algorithm \ref{DMML-TB}.

Due to the significantly smaller size of the coefficients $N^{\max}_{k,t+1}$ and $\frac{\xi^m_{k,k^{'}}}{\varphi_{k,k^{'},t}^m}$ compared to the parameter size of the training model, the overhead of transmitting these coefficients can be ignored. Additionally, the cluster head only needs to execute a simple loop to obtain the number of iterations $N^{m}_{k,t+1}$. The proposed training balance method exhibits low complexity.

\begin{figure*}[t]
	\centering
	\subfigure[]{
		\includegraphics[width=5.7cm]{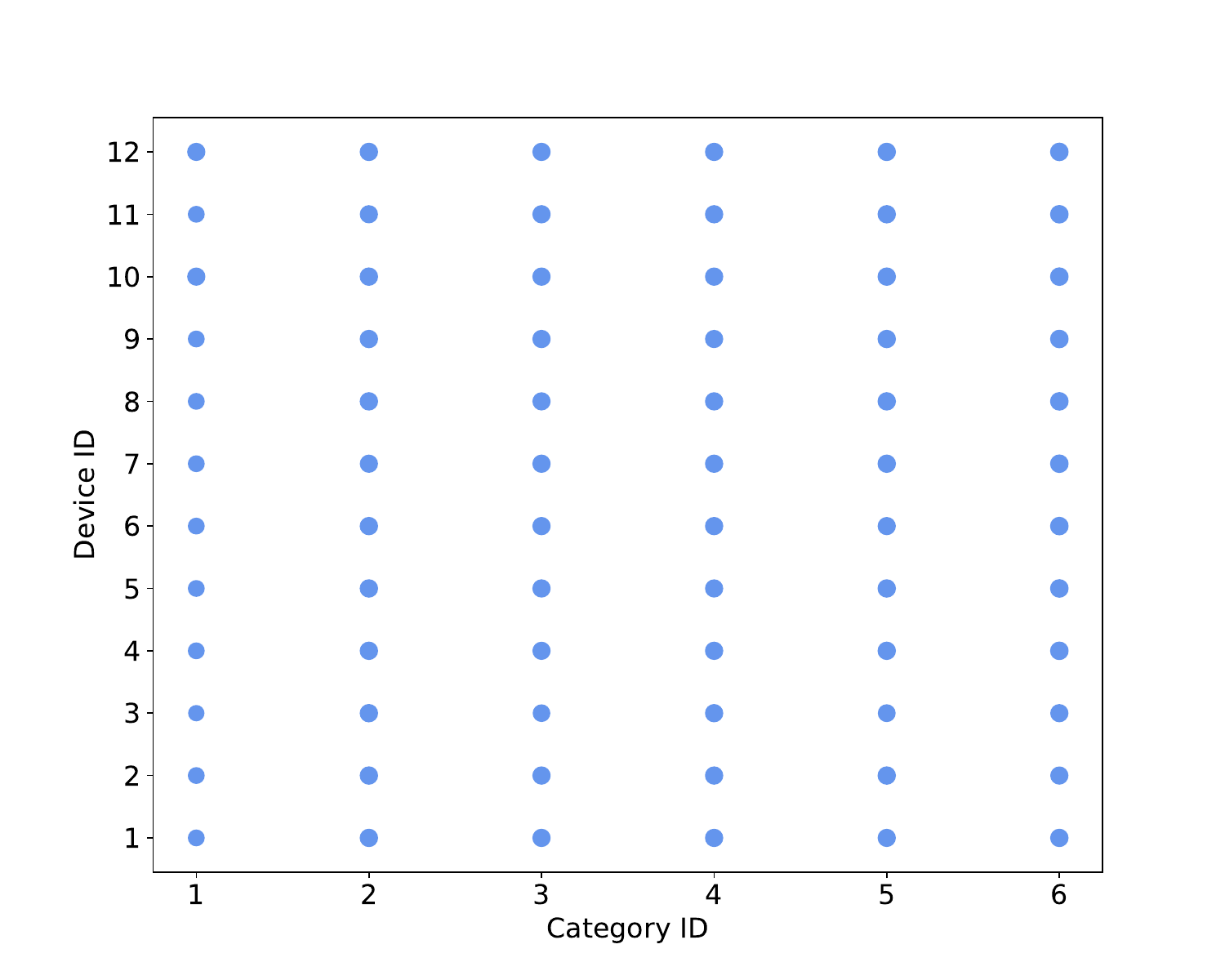}
	}
	\subfigure[]{
		\includegraphics[width=5.7cm]{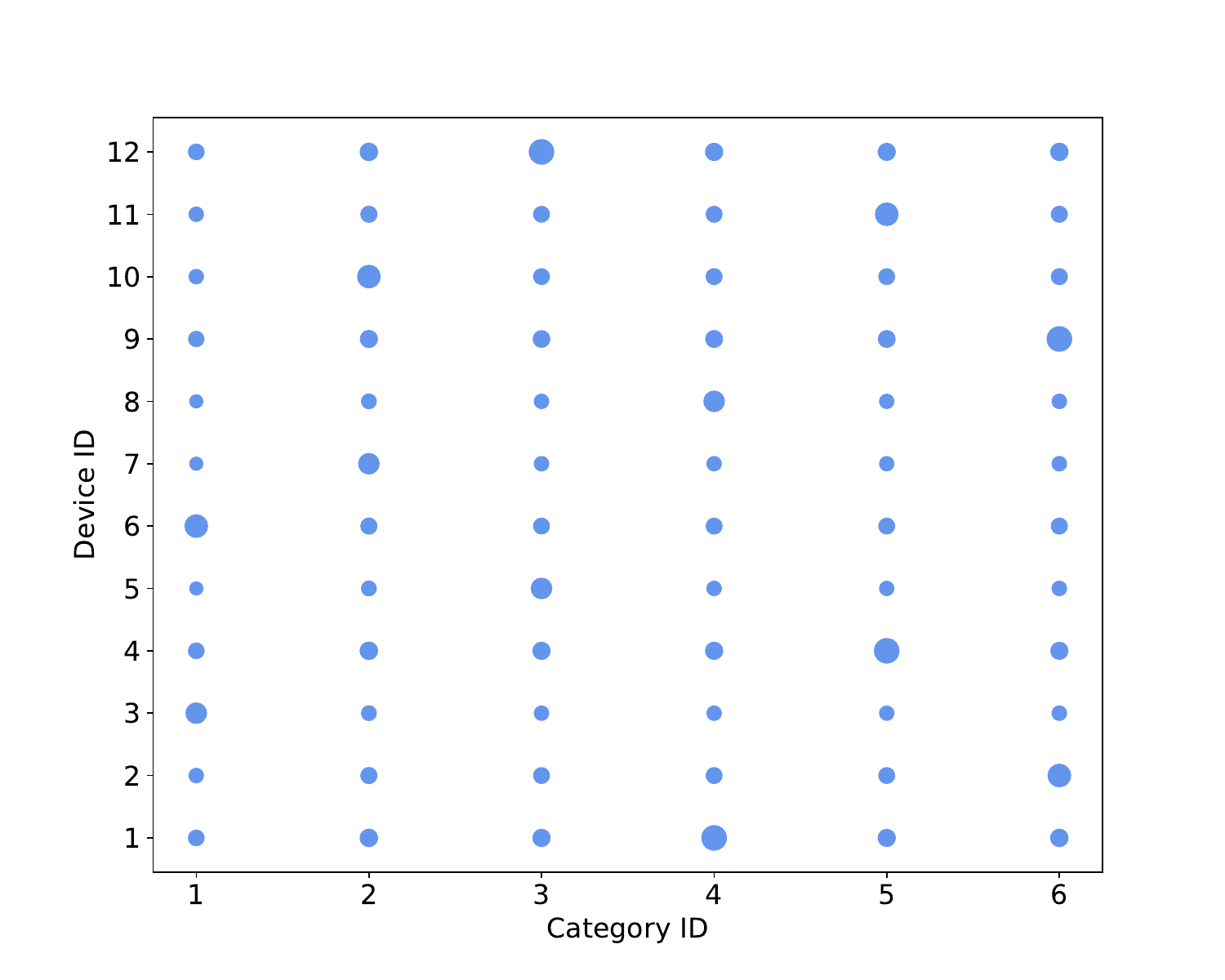}
	}
	\subfigure[]{
		\includegraphics[width=5.7cm]{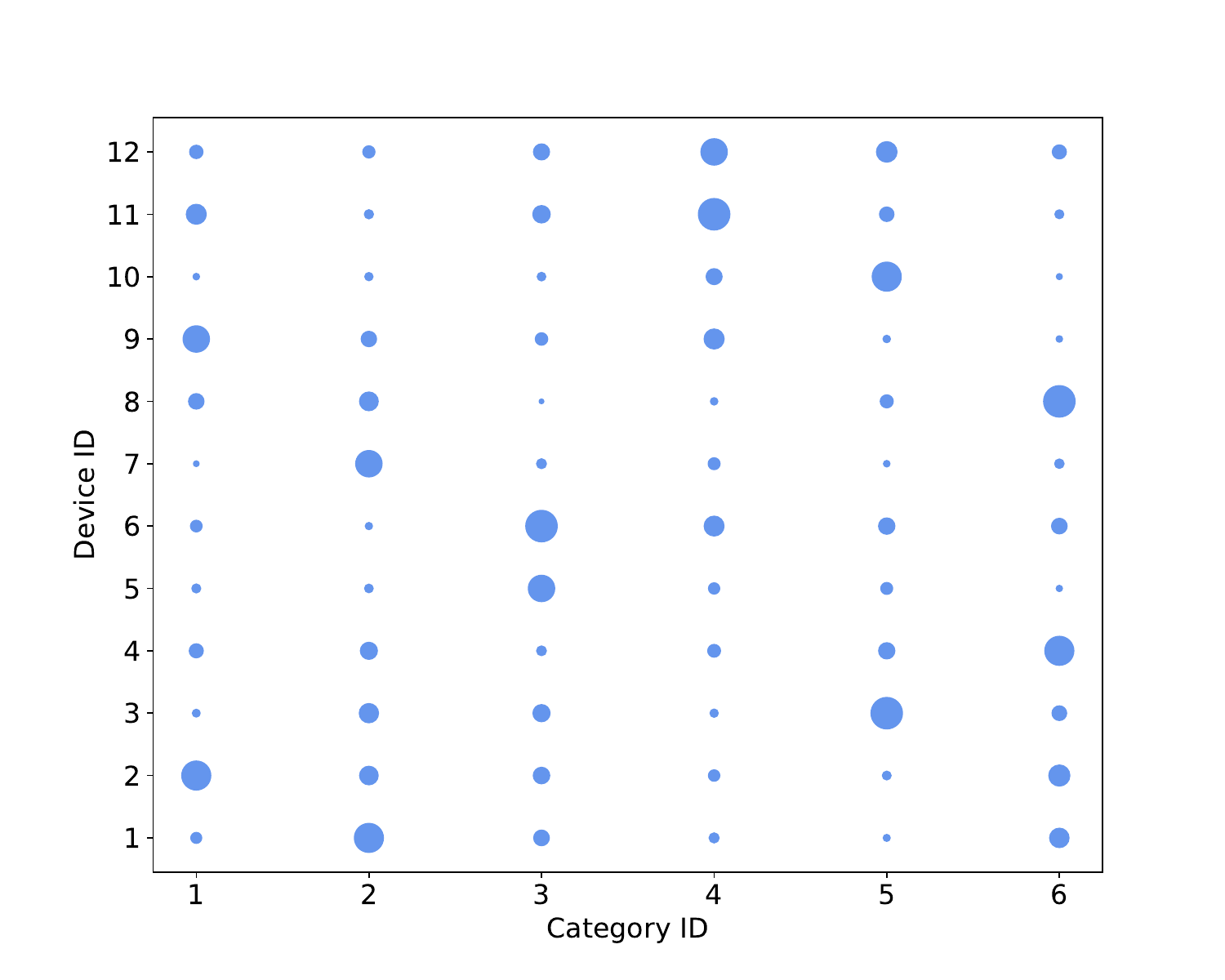}
	}
	\caption{Three different distributions of data categories. (a) $\Phi=$IID. (b) $\Phi=0.3$. (c) $\Phi=0.5$.}
	\label{non-iid}
\end{figure*}

%%%%%%%%%%%%%%%%%%%%%%%%%%%%%%%%%%%%%%%
%%%%%%%%%%%%%%%%%%%%%%%%%%%%%%%%%%%%%%%
\section{Simulation Results}
\label{sim}
In this section, we present the detail of the simulation and evaluate the performance of the proposed DMML-KD with the training balance algorithm. 

\subsection{Simulation Setup}
\begin{table}
	\centering
	\caption{Simulation parameters}
	\begin{tabular}{ll}
		\hline
		Parameter & Value \\
		\hline
		Bandwidth $\mathcal{B}$ & 500 KHz\\
		Noise power spectral density $\mathcal{N}_0$ & -140 dBm/Hz \\
		Transmission power $p_k$ & $0.1 W$\\
		CPU frequency $f_k$ & 1 GHz \\
		Number of FLOPs per cycle $e_k$ & 4 \\
		Effective capacitance coefficient $\epsilon_k$ &  $2\times 10^{-28}$ \\
		Number of global rounds $T$ & 80\\
		FLOPs specific to audio and visual $O^1,O^2$ & $\{5.73,5.82\}{\setlength\arraycolsep{0.5pt}\times}10^{10}$\\
		Parameter size specific to audio and visual $\Upsilon^1,\Upsilon^2$ & $\{8.94,8.95\}{\setlength\arraycolsep{0.5pt}\times}10^{7}$\\
		FLOPs of common classifier $\hat{O}$  & $4.92{\setlength\arraycolsep{0.5pt}\times}10^{4}$\\
		Parameter size $\Upsilon$ & $1.35{\setlength\arraycolsep{0.5pt}\times}10^{6}$\\
		FLOPs of generator $O^G_F$, $O^G$ & $\{2.7,5.4\}{\setlength\arraycolsep{0.5pt}\times}10^{6}$\\
		\hline
	\end{tabular}
	\label{sim_par}
\end{table}

\subsubsection{Simulation Environment}
A decentralized learning system with $K=12$ devices is considered in the simulation. The devices are uniformly distributed within a circular area with a diameter of 100 meters. The distance between device $k$ and $k^{'}$ is denoted as $d_{k,k^{'}}$. Device $k$ is connected to the devices within a range of 50 meters around it. The channel gain $g_{k,k^{'},t}$ follows the Rayleigh distribution with the mean $10^{-PL(d_{k,k^{'}})/20}$. The path loss is $PL(d_{k,k^{'}})(dB)=32.4+20\log_{10}(\hat{f}_k^{carrier})+20\log_{10}(d_{k,k^{'}})$ and the carrier frequency is $\hat{f}_k^{carrier}=2.6$ GHz. The simulation parameters of the considered system are summarized in Table \ref{sim_par}. 

\subsubsection{Dataset Partition}

The training dataset is CREMA-D \cite{cremad}, which is a sentiment analysis dataset. CREMA-D consists of two modalities: audio and visual. It consists of 7,442 video clips, each lasting 2-3 seconds. For the visual data, we extract one frame from each video and crop it to the size of $224\times224\times3$, which serves as the input for the feature extractor. The audio data is processed into a spectrogram with the size $257\times 188$ using the librosa \cite{librosa}. To evaluate the performance of the proposed algorithm in different scenarios, we use three different distributions of modalities, i.e., $\Gamma=1$, $\Gamma=0.5$ and $\Gamma=0$. Additionally, we apply three different distributions of data categories, i.e., $\Phi=$IID, $\Phi=0.3$ and $\Phi=0.5$. They are combined in pairs to form nine different scenarios. The specific settings for the distributions are as follows:
\begin{itemize}
	\item $\Gamma=1$: All devices possess data with both audio and visual modalities.
	
	\item $\Gamma=0.5$: Half of the devices have data with both audio and visual modalities, while the other devices have data with only one of the modalities, either audio or visual.
	
	\item $\Gamma=0$: All devices have data with only one of the modalities, either audio or visual.
	
	\item $\Phi=$IID: The data on each device is randomly selected from the entire dataset, as shown in Fig.\ref{non-iid} (a).
	
	\item $\Phi=0.3$: 30\% data on each device belongs to one category, while the remaining data is randomly composed of the other category data, as shown in Fig.\ref{non-iid} (b).
	
	\item $\Phi=0.5$: 50\% data on each device belongs to one category, while the remaining data is randomly composed of other category data, as shown in Fig.\ref{non-iid} (c).
	
\end{itemize}

We employ ResNet-18 \cite{resnet} to extract features from both the audio and visual data. The length of the extracted features is $L=512$, and the length $\hat{L}=256$. The common classifier and the modality-specific classifiers are the fully connected layer with 256 input nodes and $C=6$ output nodes. The FLOPs and the parameter size of the training network are summarized in Table \ref{sim_par}.

\begin{figure}[t]
	\centering
	\includegraphics[width=7cm]{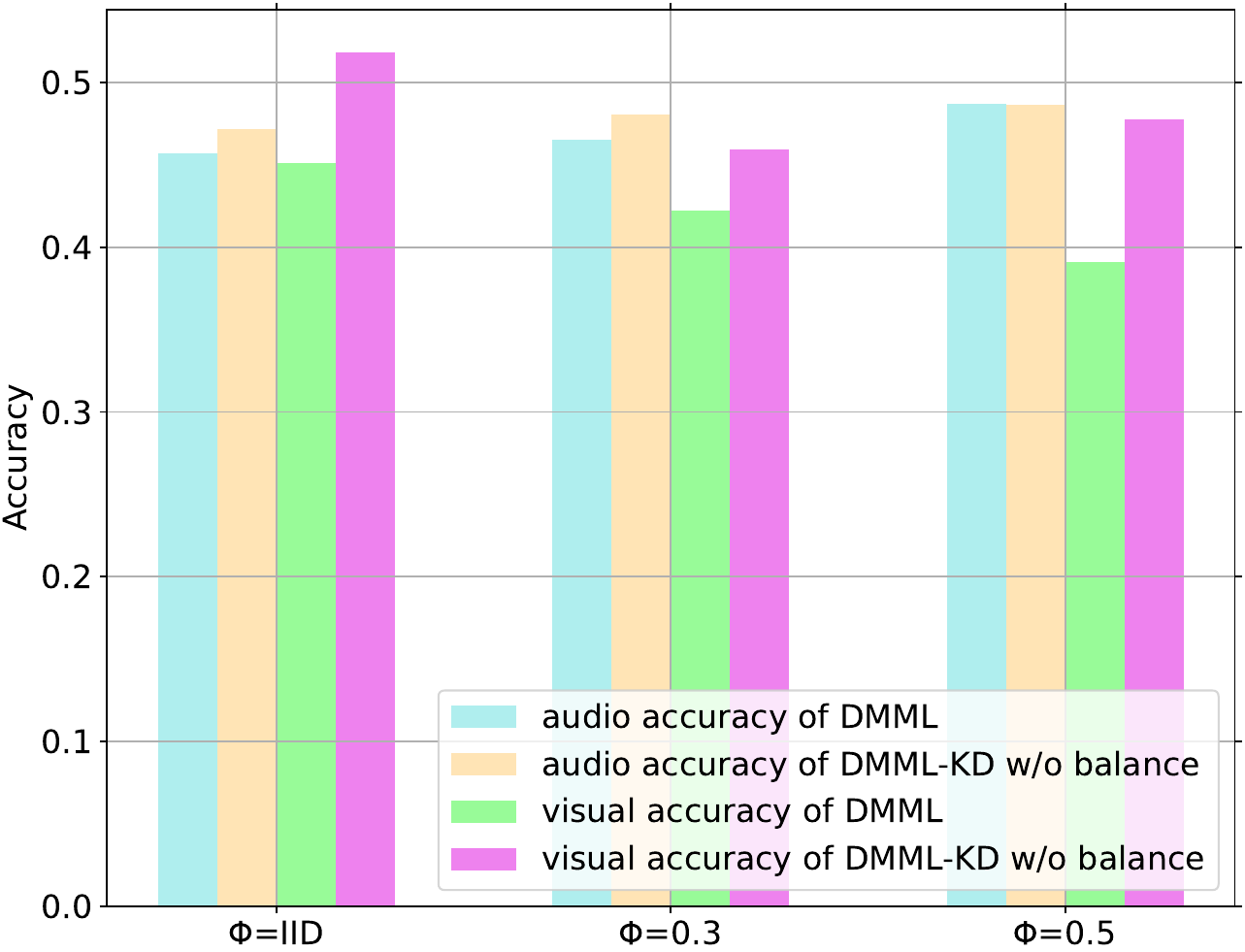}
	\caption{Audio and visual accuracy with $\Gamma=0$.}
	\label{gamma0}
\end{figure}
\subsubsection{Hyper-parameters of DMML-KD}
The generator is a two-layer fully connected network with the nodes [70, 512, 256] sequentially. The learning rate of the multi-modal network and the generator are $5\times 10^{-4}, 1\times 10^{-4}$, respectively. The number of iterations for the generator is 50 in each round. The hyper-parameters $\alpha_1,\alpha_2,\alpha_3,\alpha_4, \beta$ in the loss functions are all set as 1. The losses are optimized with the Adam optimizer \cite{adam}. The aggregation coefficients are obtained with the Metropolis-Hastings weights \cite{MH_weight}.

\subsection{Performance Evaluation}
\begin{figure*}[t]
	\centering
	\subfigure[]{
		\includegraphics[width=7cm]{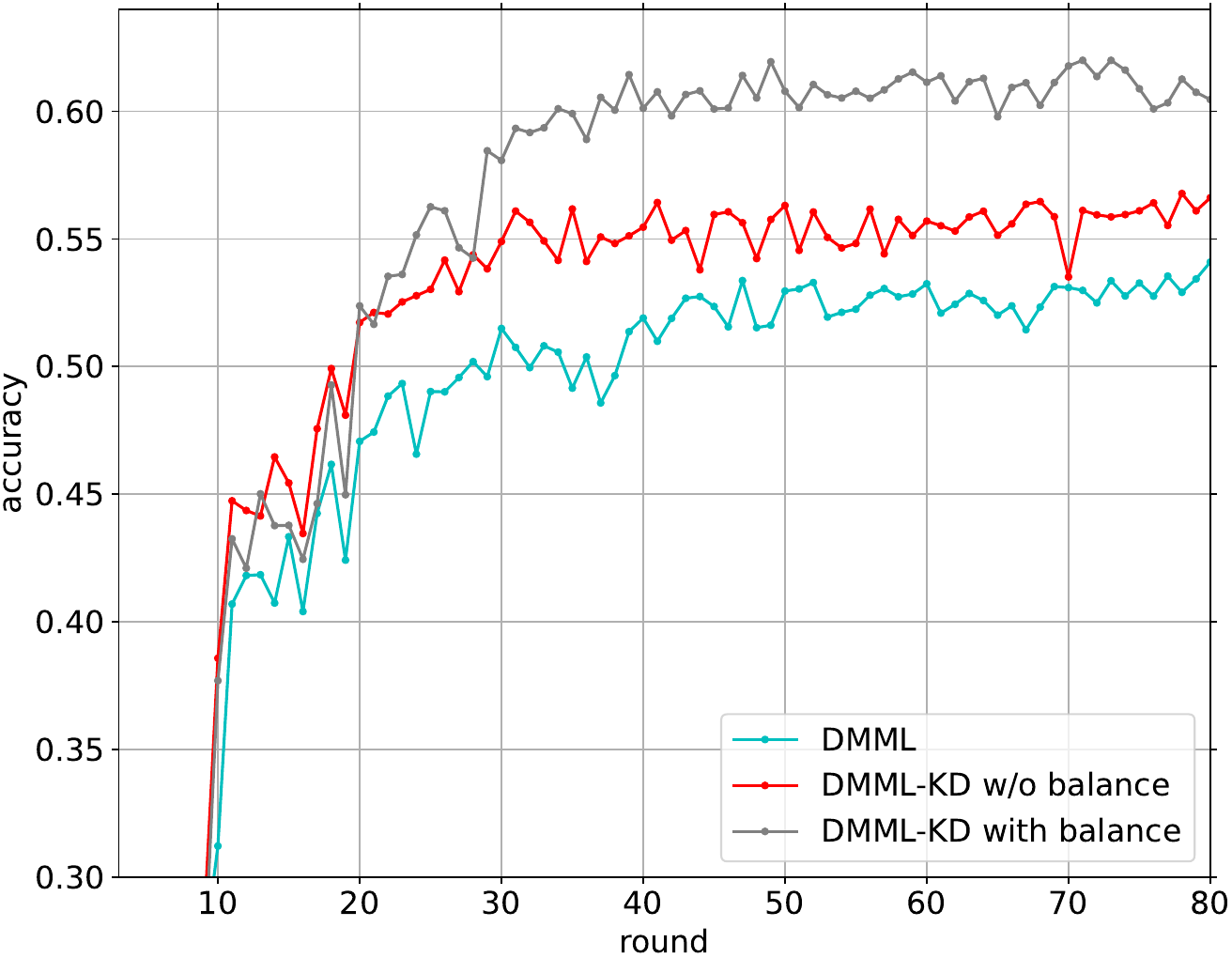}
	}
	\subfigure[]{
		\includegraphics[width=7cm]{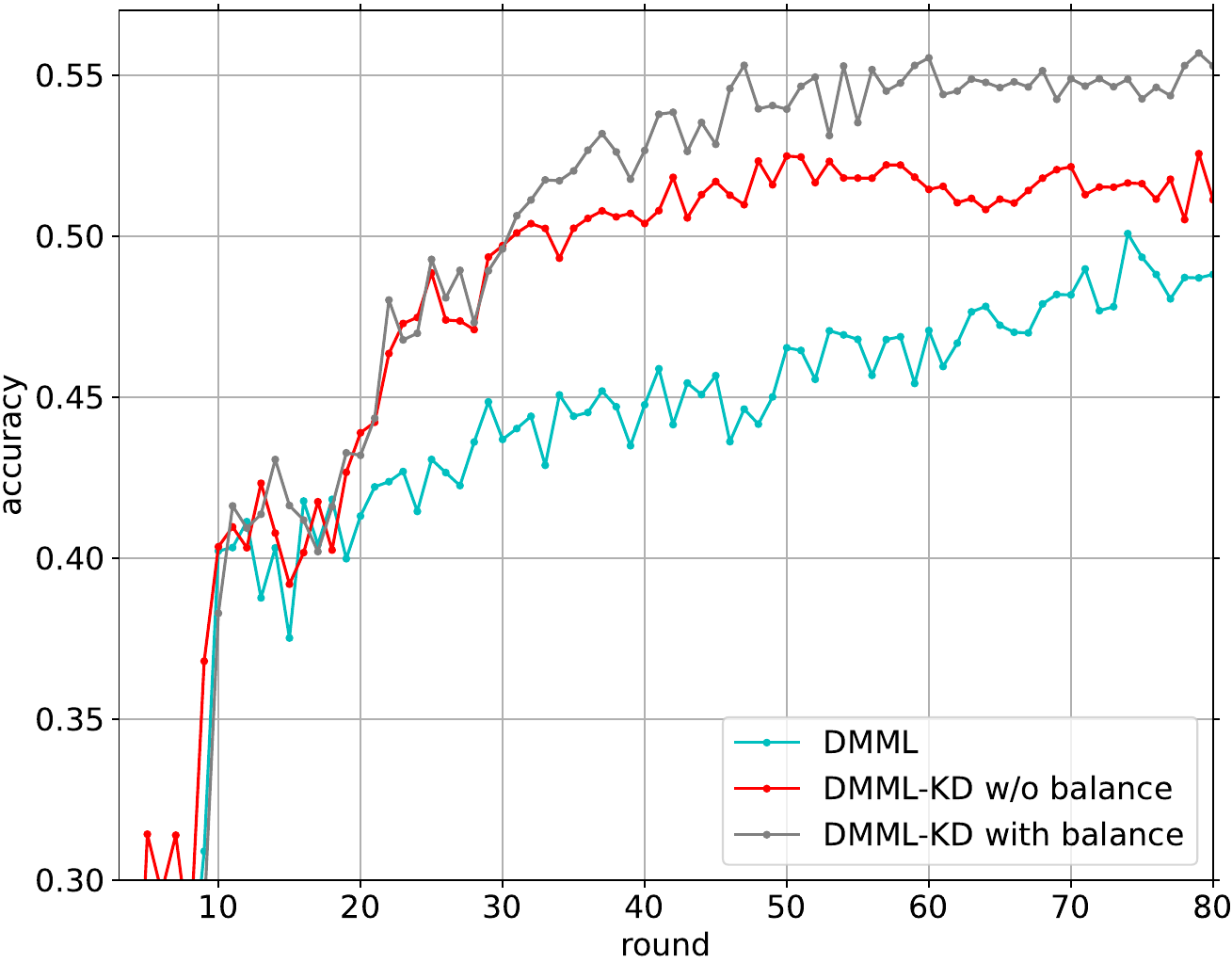}
	}
	\caption{Test accuracy with $\Phi=$IID and (a) $\Gamma=1$, (b) $\Gamma=0.5$.}
	\label{phiiid}
\end{figure*}

\begin{figure*}[t]
	\centering
	\subfigure[]{
		\includegraphics[width=7cm]{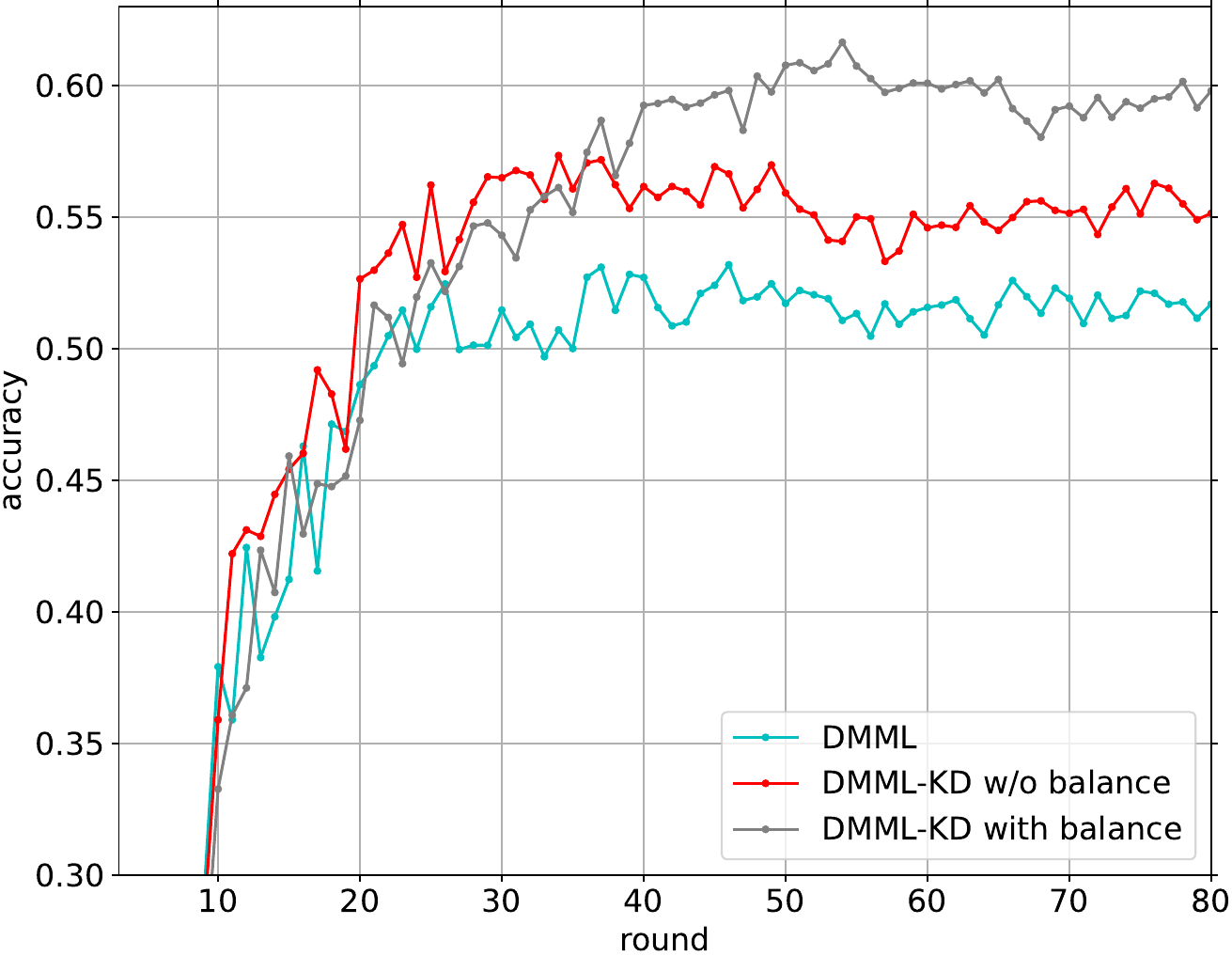}
	}
	\subfigure[]{
		\includegraphics[width=7cm]{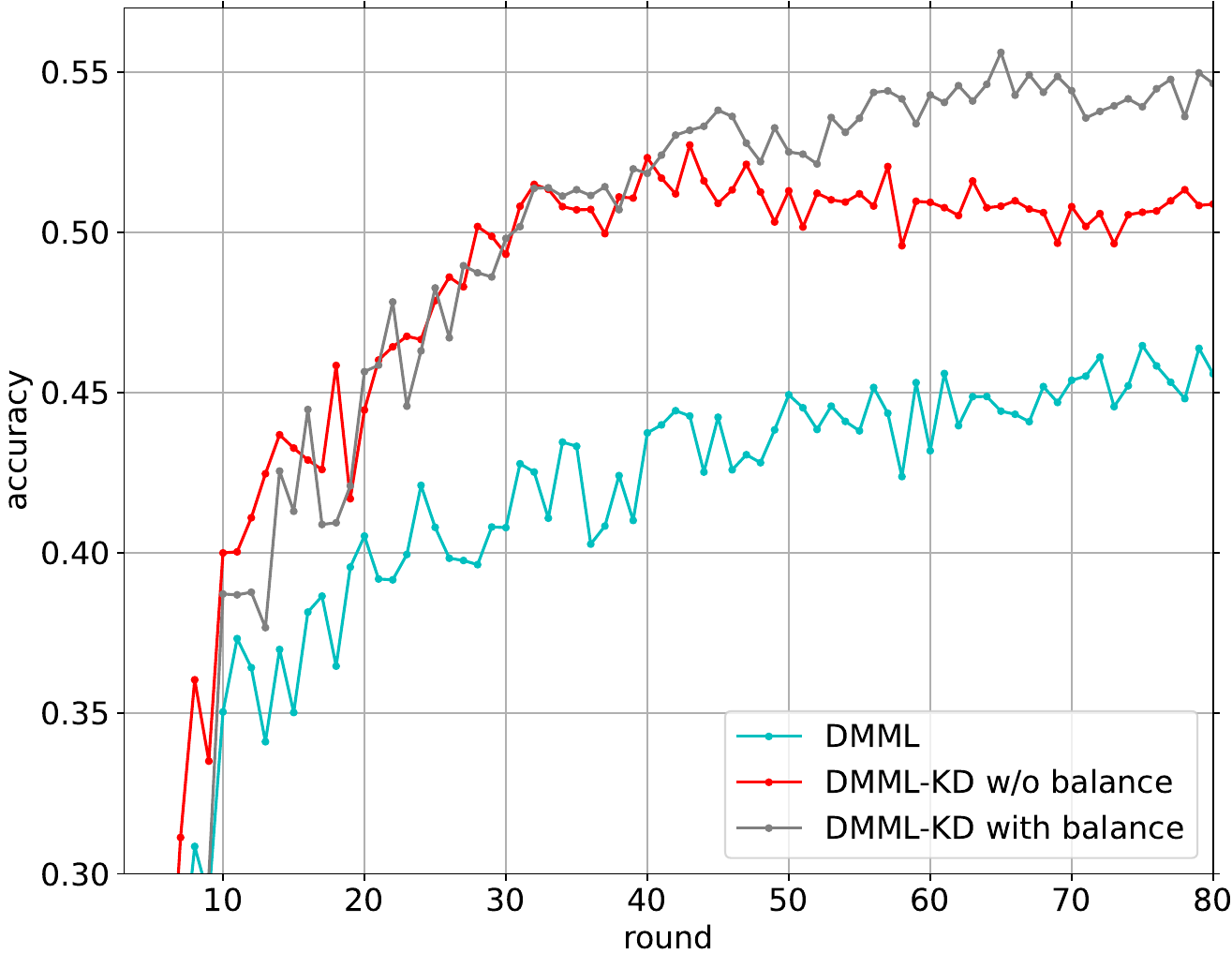}
	}
	\caption{Test accuracy with $\Phi=$0.3 and (a) $\Gamma=1$, (b) $\Gamma=0.5$.}
	\label{phi03}
\end{figure*}

\begin{figure*}[t]
	\centering
	\subfigure[]{
		\includegraphics[width=7cm]{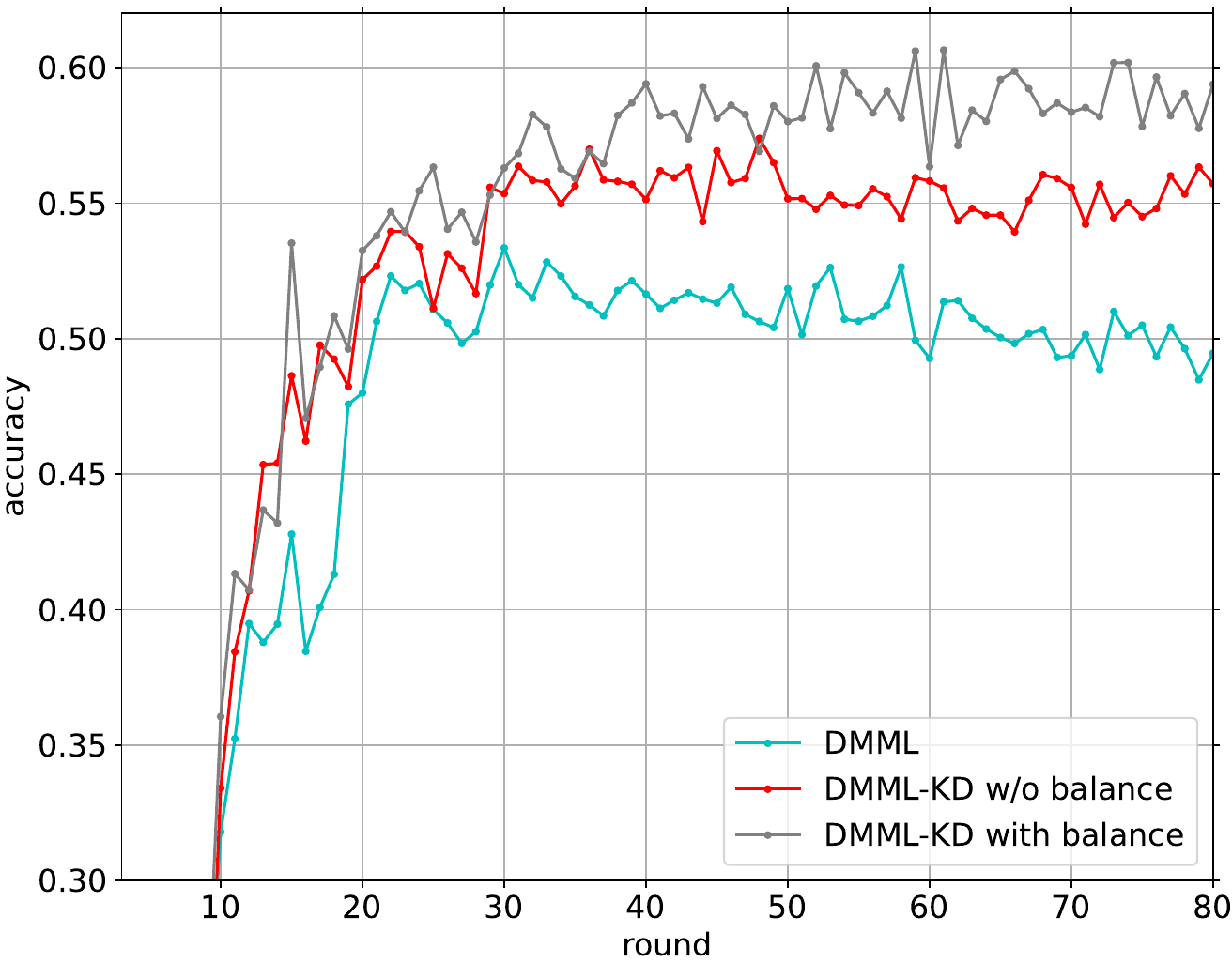}
	}
	\subfigure[]{
		\includegraphics[width=7cm]{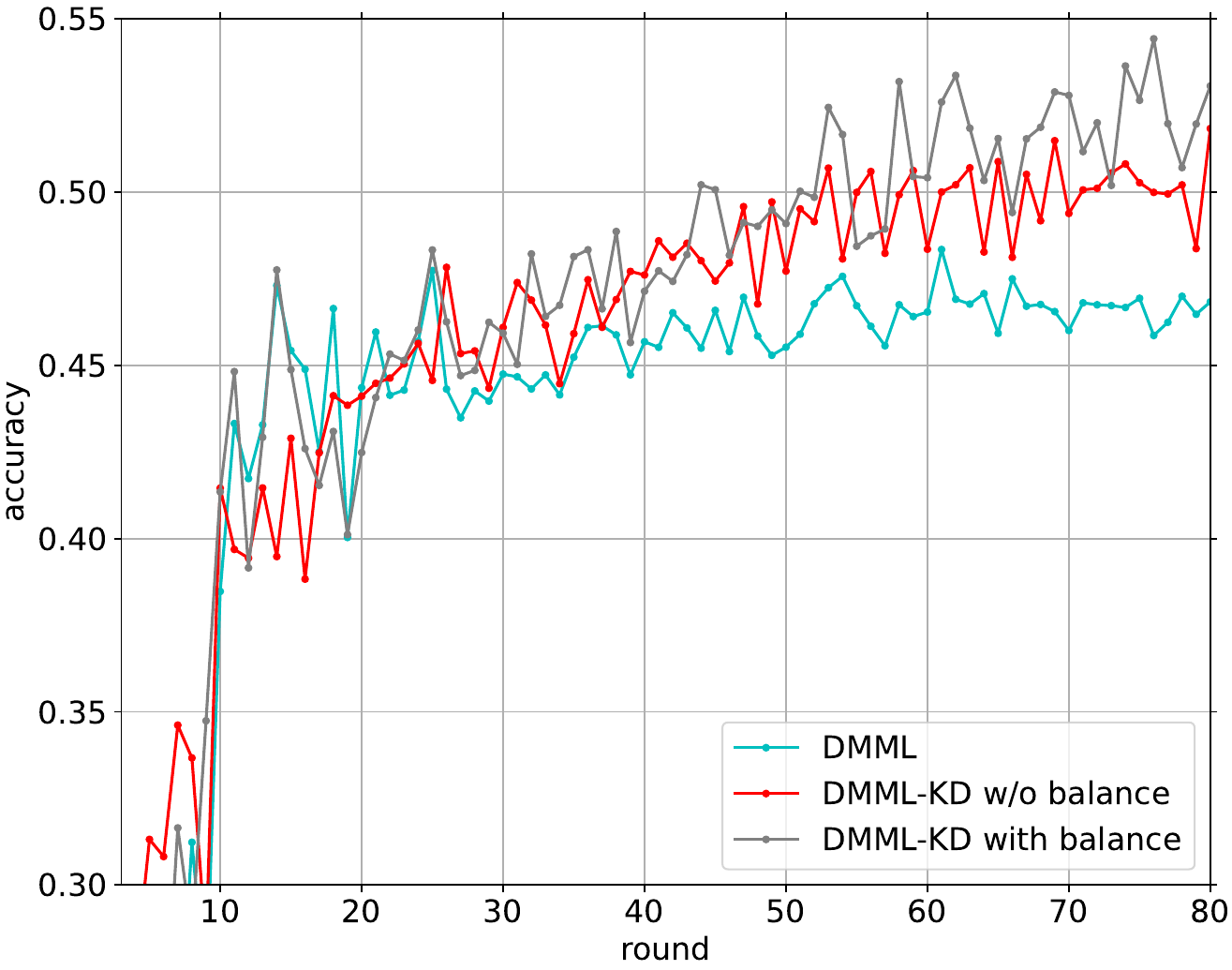}
	}
	\caption{Test accuracy with  $\Phi=$0.5 and (a) $\Gamma=1$, (b) $\Gamma=0.5$.}
	\label{phi05}
\end{figure*}

\begin{figure}[t]
	\centering
	\includegraphics[width=7cm]{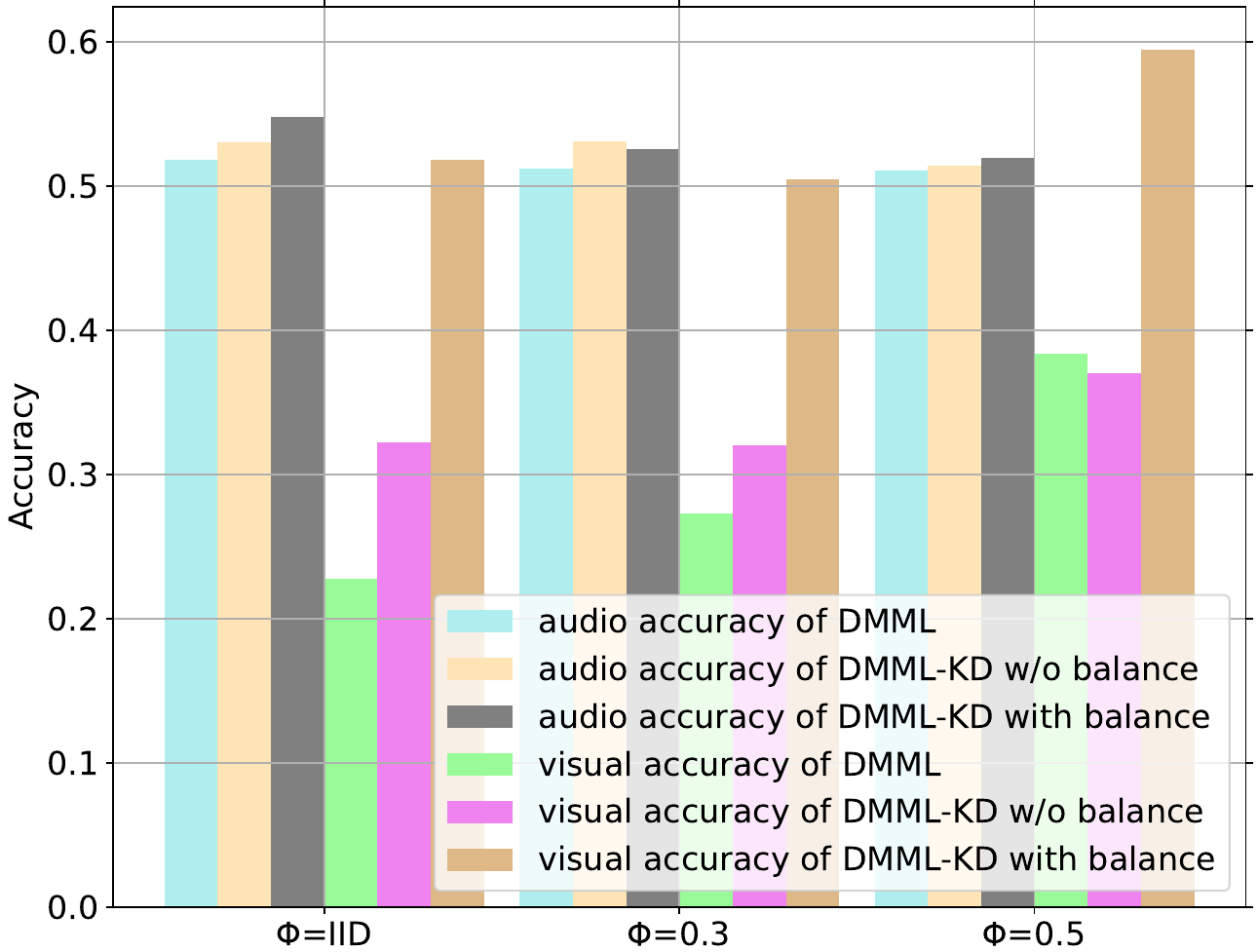}
	\caption{Audio and visual accuracy with $\Gamma=1$.}
	\label{gamma1}
\end{figure}

\begin{figure}[t]
	\centering
	\includegraphics[width=7cm]{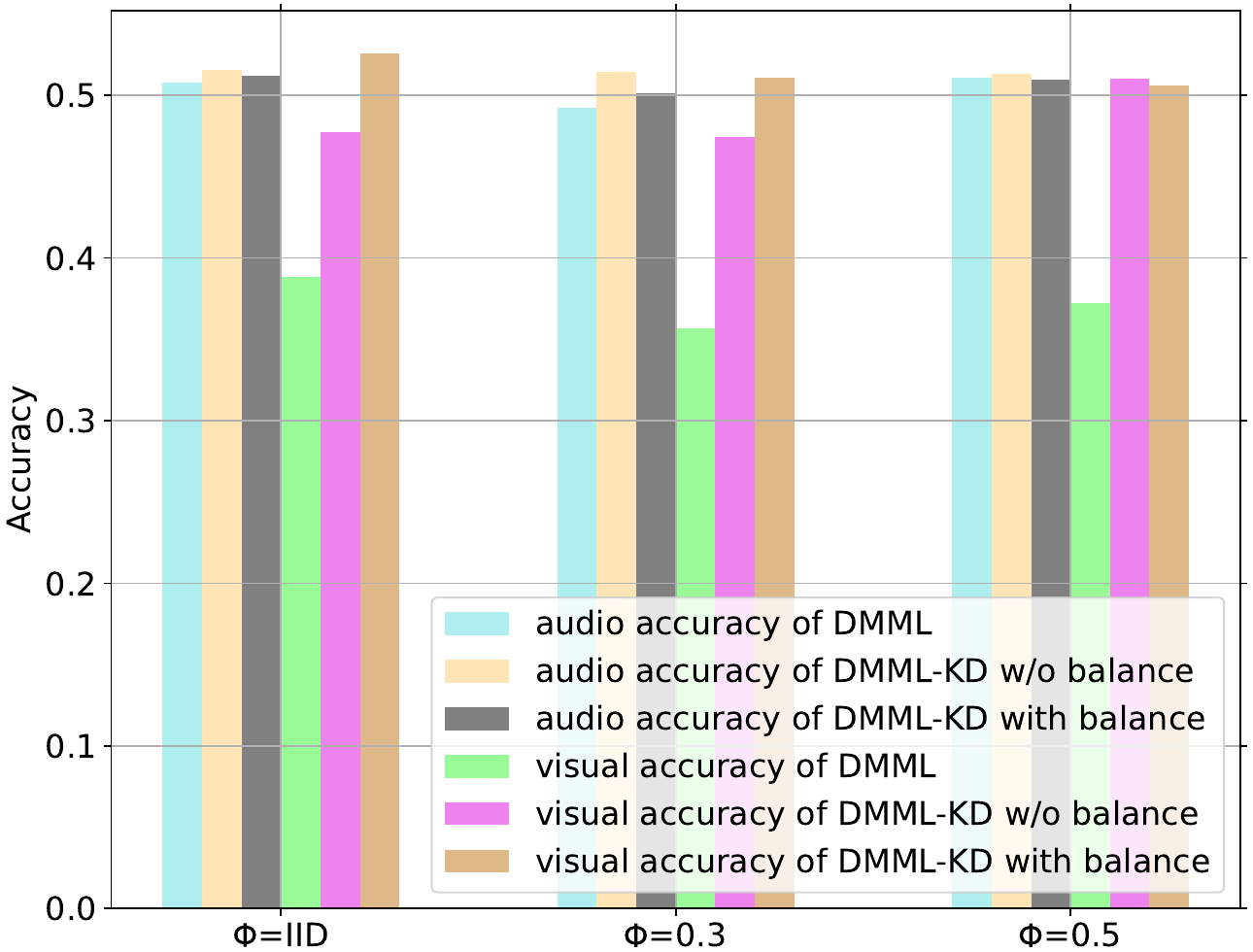}
	\caption{Audio and visual accuracy with $\Gamma=0.5$.}
	\label{gamma05}
\end{figure}

The test accuracies of DMML and DMML-KD under different data partitions are shown in Fig. \ref{gamma0}, Fig. \ref{phiiid}, Fig. \ref{phi03} and Fig. \ref{phi05}. Specifically, in the case of $\Gamma=0$, the model trained on devices is single-modal, and there is no issue of imbalanced training across modalities. Therefore, training balancing is not performed in the case of $\Gamma=0$. For the cases of $\Gamma=1$ and $\Gamma=0.5$, the test accuracies of different devices are averaged to obtain the curves shown in Fig. \ref{phiiid}, Fig. \ref{phi03} and Fig. \ref{phi05}. In the case of $\Gamma=0$, the test accuracies of devices with only the visual modality and only the audio modality are separately averaged to obtain the visual accuracy and audio accuracy, as shown in Fig. \ref{gamma0}. To estimate the visual accuracy and audio accuracy of the multi-modal network in the cases of $\Gamma=1$ and $\Gamma=0.5$, we use the following term \cite{OGM} as the approximated prediction of modality $m\in \{1,2\}$,
\begin{align}
	\bm{c}^m_{k,d}=\hat{\bm{w}}_{k,t}^n\hat{\bm{h}}_{k,d}^{m}+\check{\bm{w}}_{k,t}^{n,m}\check{\bm{h}}_{k,d}^{m} +\frac{\bm{b}_{k,t}^{n}}{2}.
\end{align}
The estimated accuracies are shown in Fig. \ref{gamma1} and Fig. \ref{gamma05}.
\subsubsection{Knowledge Distillation}
As shown in Fig. \ref{phiiid}, Fig. \ref{phi03} and Fig. \ref{phi05}, knowledge distillation can effectively improve the accuracy. In the case of $\Gamma=0$, the single-modal model trained with DMML cannot learn the correlated features of the other modality. As shown in Fig. \ref{gamma0}, after applying knowledge distillation, the visual and audio accuracy are both improved, which indicates that knowledge distillation can transfer feature knowledge of the other modality. 

Moreover, as shown in Fig. \ref{phiiid} (a), Fig. \ref{phi03} (a) and Fig. \ref{phi05} (a), the performance gain between DMML-KD without balance and DMML is 2.69\% in the case of $\Phi=$IID. In contrast, the performance gains for the cases of $\Phi=0.3$ and $\Phi=0.5$ are 4.15\% and 4.04\%, respectively. This is because the performance of DMML-KD is almost unchanged in the presence of non-IID data, indicating that DMML-KD can deal with the non-IID data.

\subsubsection{Training Balance}
As shown in Fig. \ref{gamma0}, when training a single-modality network with DMML, both the visual and audio feature extractor are trained effectively. The visual accuracy can approach or even exceed the audio accuracy. However, the visual accuracy is significanty lower than that of the audio accuracy when joint training the multi-modal network, as shown in Fig. \ref{gamma1}. The visual accuracy in Fig. \ref{gamma05} is higher than that in  Fig. \ref{gamma1}, which is because a portion of devices train the visual-only network under $\Gamma=0.5$. These results indicate that training a multi-modal network jointly may lead to the insufficient training of the visual feature extractor.

As shown in Fig. \ref{phiiid}, Fig. \ref{phi03} and Fig. \ref{phi05}, compared to DMML-KD without balance, the proposed training balance method can further improve the accuracy. As shown in Fig. \ref{gamma1} and Fig. \ref{gamma05}, although DMML-KD without balance can improve the visual and audio accuracy compared to DMML, the training of the visual modality is still not sufficiently compared to the audio modality. The visual accuracy is improved by the proposed training balance method. Besides, it can be observed from Fig. \ref{gamma1} that different category distributions can impact the visual accuracy obtained with DMML. In contrast, our proposed DMML-KD with training balance can achieve comparable or even higher visual accuracy than the audio accuracy across different non-IID scenarios.

Apart from the performance gain, the proposed training balance method can also reduce the energy consumption of devices, as shown in Table \ref{energy_save}. Compared with DMML, training and transmitting the generator in DMML-KD add some computational load on devices. However, the required FLOPs of training the generator is much smaller than that of the feature extraction networks. The energy saved in devices by reducing the number of iterations exceeds the energy consumption added by training the generator.

\begin{table}
	\centering
	\caption{Averaged energy consumption per device with $\Gamma=1$}
	\begin{tabular}{ccc}
		\hline
		 Distribution & DMML ($\times 10^4$ J) & DMML-KD with balance($\times 10^4$ J) \\
		\hline
		 $\Phi=$IID&1.737 &1.652   \\
		 $\Phi=0.3$&1.749 &1.645  \\
		 $\Phi=0.5$&1.749 &1.671 \\
		\hline
	\end{tabular}
	\label{energy_save}
\end{table}

%%%%%%%%%%%%%%%%%%%%%%%%%%%%%%%%%%%%%%%
\section{Conclusion}
\label{con}
In this paper, we propose the DMML with knowledge distillation to enhance the training performance in the presence of non-IID data and modality heterogeneity over wireless networks. The proposed method utilizes a generator to learn the global conditional distribution of modality-common features and guide the modality-common features of different devices towards the same distribution. Furthermore, we design a strategy to reduce the number of local iterations for rapidly converging modalities, addressing the issue of training imbalance. The optimization of the number of local iterations for different modalities on each device is achieved based on the proposed performance metric. Experimental results show that the proposed DMML-KD framework with training balance significantly improves the training performance of DMML.

%\begin{appendices}
%\section{Proof of Lemma 1}
%\label{lemma1}
%
%\end{appendices}

\bibliography{reference}

\end{document}